%% file: ms.tex
\documentclass[twocolumn]{aastex631}

\newcommand{\R}{\textcolor{red}}

\usepackage{amsmath}
 
\shorttitle{The Mid-Infrared Transmission Spectrum of TOI-270~d}
\shortauthors{Holmberg et al.}

\graphicspath{{./}{figures/}}

\begin{document}

\title{The Mid-Infrared Transmission Spectrum of the Temperate Sub-Neptune TOI-270~d}

\author[0000-0002-0931-735X]{Måns Holmberg}
\affiliation{Space Telescope Science Institute, 3700 San Martin Drive, Baltimore, MD 21218, USA}
\email{Correspondence: nmadhu@ast.cam.ac.uk, mholmberg@stsci.edu}

\author[0000-0002-4869-000X]{Nikku Madhusudhan}
\affiliation{Institute of Astronomy, University of Cambridge, Madingley Road, Cambridge CB3 0HA, UK}

\author[0000-0002-0931-735X]{Martin Binet}
\affiliation{Institute of Astronomy, University of Cambridge, Madingley Road, Cambridge CB3 0HA, UK}

\author[0000-0002-2705-5402]{Subhajit Sarkar} \affiliation{School of Physics and Astronomy, Cardiff University, The Parade, Cardiff CF24 3AA, UK}

\author[0009-0001-9140-3299]{Frances E. Rigby} \affiliation{Institute of Astronomy, University of Cambridge, Madingley Road, Cambridge CB3 0HA, UK}

\author[0000-0002-8837-0035]{Julianne I. Moses} \affiliation{Space Science Institute, Boulder, CO 80301, USA}

\begin{abstract}

Observations of temperate sub-Neptunes with JWST have ushered in a new era for atmospheric characterization of small exoplanets. In particular, the MIRI instrument provides a unique opportunity to search for molecules that are not easily accessible in the near-infrared, as demonstrated by recent mid-infrared observations of K2-18~b. In this work, we present the first mid-infrared transmission spectrum of TOI-270~d observed using the JWST MIRI LRS (5-12~$\mu$m) instrument. By leveraging archival MIRI LRS data, we establish a new empirical relation between the detector settling timescale and the flux, which helps accurately model the spectral light curves and improve the precision of the transmission spectrum. We find that there is notable evidence of molecular features in the MIRI transmission spectrum of TOI-270~d, favouring the presence of atmospheric absorption at $\ln B$~=~2.8-5.3 when comparing physically plausible atmospheric models with and without molecular line absorption. The data show excess absorption beyond what could be attributed to CH$_4$ and CO$_2$ detected previously, in line with recent near-infrared results. Through an agnostic search for 203 species, we identify several candidate trace molecules, most of which are complex molecules, evaluate their physical plausibility, and compare them against inferences from near-infrared observations. We also compare the MIRI spectrum of TOI-270~d to that of K2-18~b and find that random or systematic noise is unlikely to explain these observations. Future follow-up observations are necessary to definitively identify the additional absorbers beyond CH$_4$ and CO$_2$. These observations demonstrate the unique capability of JWST MIRI for atmospheric characterisation of temperate sub-Neptunes.

\end{abstract}

\section{Introduction} \label{sec:intro}

Temperate sub-Neptunes ($T_\mathrm{eq}\lesssim400$ K, $R_{\mathrm{p}} \sim 1–4~$R$_\oplus$) represent a new frontier in exoplanet science with JWST. These planets have bulk properties consistent with a wide range of internal structures and lack solar system analogues to guide interpretation. Recent near-infrared (NIR) observations have begun to unveil their atmospheric compositions, starting with the first detections of carbon-bearing molecules in the atmospheres of K2-18~b \citep{Madhusudhan2023b, hu_water-rich_2025} and TOI-270~d \citep{Holmberg2024, Benneke2024, Felix2025, Constantinou2025}. In both cases, CH$_4$ and CO$_2$ were robustly detected, while DMS was tentatively inferred in K2-18 b and H$_2$O, CS$_2$, and other sulfur-bearing species were suggested in TOI-270 d. Using MIRI LRS \citep{Kendrew2015}, \cite{Madhusudhan2025} recently reported possible detections of DMS and/or DMDS -- potential biosignature gases \citep{pilcher2003, domagal-goldman2011, seager2013b} -- in K2-18~b, demonstrating the capability of mid-infrared (MIR) spectroscopy with JWST to probe complex molecules in small exoplanet atmospheres. However, additional MIR observations are essential to assess the robustness of these findings.

The observation of a MIR spectrum of a habitable-zone planet marks a significant advancement in our ability to study complex molecules \citep{Madhusudhan2025}, which are generally easier to resolve at longer wavelengths. Yet, MIR spectroscopy faces challenges due to its intrinsically lower signal-to-noise ratio (SNR) relative to NIR data. The interpretation of the K2-18~b MIRI spectrum has been debated on multiple levels. While some argue that the spectrum is featureless \citep{Taylor2025}, others argue that the features are too large and inconsistent with the NIR data \citep{Luque2025,Yassin2025}, resulting in contradictory conclusions. If the features are indeed real, they could represent DMS or some other complex molecule \citep{Madhusudhan2025, welbanks_challenges_2025, pica-ciamarra_systematic_2025}. If not, they might result from potential systematic uncertainties \citep{Stevenson2025}. More observations are needed to robustly distinguish between these different interpretations, through repeat observations of K2-18~b and/or through independent observations of other temperate sub-Neptunes. Here we present the latter: the first MIR transmission spectrum of TOI-270~d obtained with MIRI LRS.

TOI-270~d orbits a quiet M3 star with a period of 11.4 days \citep{Gunther2019}. It has a mass of $M_\mathrm{p}=4.78\pm0.43~\mathrm{M_\oplus}$ \citep{VanEylen2021} and a radius of $R_\mathrm{p}= 2.19\pm0.07~\mathrm{R_\oplus}$ \citep{Mikal-Evans2023}, consistent with a range of interior compositions spanning gas dwarf, mini-Neptune, and hycean world scenarios \citep[e.g.][]{VanEylen2021,Madhusudhan2021,Rigby2024a}. Hycean planets, with liquid water oceans beneath thin H$_2$-rich atmospheres, have been proposed as potential hosts of habitable conditions \citep{Madhusudhan2021}. With an equilibrium temperature of 387~K assuming $A_\mathrm{B}=0$ (or 326~K for $A_\mathrm{B}=0.5$), TOI-270~d is warmer than K2-18~b, making a stable liquid ocean surface less feasible given the increased potential for runaway greenhouse and convective inhibition \citep[e.g.][]{Leconte2024}. 

The nature of TOI-270~d remains debated \citep[][]{Holmberg2024, Benneke2024, Felix2025, Glein2025, Nixon2025, Rigby2026}. In the case of K2-18~b, the abundances of CH$_4$ and CO$_2$ and non-detections of NH$_3$ and CO derived from its JWST transmission spectrum in the NIR \citep{Madhusudhan2023} were found to be consistent with prior predictions for a Hycean world \citep{Hu2021, Tsai2021, Madhusudhan2023}. Subsequent studies debated other possibilities for its internal structure, including mini-Neptune \citep[e.g.][]{Wogan2024, Cooke2024} and gas dwarf scenarios \citep{Shorttle2024, Rigby2024b}. Most recently, high-precision abundance estimates with repeated NIR observations robustly confirmed the previous abundance estimates and support a water-rich interior in the planet \citep{hu_water-rich_2025}. A similar molecular pattern was also reported for TOI-270~d \citep{Holmberg2024}, though its higher temperature suggests that it may be too warm to sustain a habitable liquid water ocean, especially on the dayside. Alternatively, the higher mean molecular weight atmosphere inferred by \cite{Benneke2024} potentially supports a mixed-envelope sub-Neptune scenario, comprising a high-metallicity envelope over a rocky core. Refining the atmospheric abundances and understanding possible atmosphere–interior interactions are crucial for constraining the planet's true nature.

Beyond the abundances of prominent molecules, the NIR observations of TOI-270~d also indicate the possibility of excess absorption due to the other trace species in the atmosphere. Multiple recent studies have explored different molecules to explain the excess absorption, from complex hydrocarbons to sulfur-bearing species \citep{Felix2025, Constantinou2025}. Similar inferences of excess absorption have also been made in both the NIR and MIR for other temperate sub-Neptunes, K2-18~b \citep{Madhusudhan2023b, Madhusudhan2025, pica-ciamarra_systematic_2025} and TOI-732 c \citep{Rigby_Madhusudhan_2025}. In this work, we present the first JWST MIRI LRS transmission spectrum of TOI-270~d, complementing previous NIR observations. 

In what follows, we describe the observations, light-curve analysis, and data reduction in Section~\ref{sec:obs}, and the atmospheric retrievals and results in Section~\ref{sec:retrieval}. We summarize and discuss our findings in Section~\ref{sec:discussion}.

\section{Observations} 
\label{sec:obs}

We observed one transit of TOI-270~d using MIRI Low Resolution Spectrograph (LRS) \citep{Kendrew2015} in slitless mode as part of JWST GO program 3557 (PI: N. Madhusudhan). The science exposure took place on July 26th 2024 between  01:29:06.762 and 05:46:53.879 UTC, a duration of 4.3 hours (2$\times$ the transit duration). We obtained a total of 5403 integrations, with 17 groups per integration. The SLITLESSPRISM subarray was used with the FASTR1 read out mode.
No high-gain antenna movement was reported during the observation. 
We processed the data with two independent reductions using the \texttt{JExoRES} and \texttt{JexoPipe} pipelines.
In the sections below, we describe the data reduction process and light curve fitting for each pipeline.  Under \texttt{JExoRES}, we also describe a new approach to determining the detector settling timescale as a function of flux.  Ultimately, we obtain MIRI LRS transmission spectra for TOI-270~d that cover the wavelength range $\sim$5–12~$\mu$m.  

For our atmospheric retrieval analyses, we also use the near-infrared spectrum of TOI-270 d from GO program 4098 (PI: B. Benneke), as reported in \cite{Holmberg2024} and \cite{Constantinou2025}. These observations used NIRISS SOSS and NIRSpec G395H, covering 0.6–5.2~$\mu$m.

\subsection{JExoRES} \label{sec:reduction}

 We employ the latest version of \texttt{JExoRES} which has previously been applied to NIRISS, NIRSpec and MIRI observations  \citep{Madhusudhan2023b,Holmberg2024, Madhusudhan2025}. Below, we describe the data reduction, characterization of the settling timescale, and light curve fitting.

\subsubsection{Data reduction} \label{sec:reduction}

To start, we perform Stages 1–2,
using the JWST Science Calibration Pipeline \citep{Bushouse2020} to conduct the detector calibration and the detector ramp fitting. As is standard for MIRI LRS observations in Stage~1, we first perform the data quality initialization, electromagnetic interference (EMI) correction, saturation flagging, first and last frame flagging, linearity correction, reset switch charge decay (RSCD) masking, dark current subtraction, and ramp fitting. In Stage~2, we apply the flat-field correction.

Because our integrations contain only 17 groups, the standard RSCD masking -- which removes the first five groups -- eliminates a significant fraction of the ramp and reduces the SNR. We therefore also perform the reduction without RSCD masking. This no-RSCD reduction yields higher SNR and is adopted as our default \texttt{JExoRES} dataset for the main analysis. However, we consider both reductions for the canonical retrievals in Section~\ref{sec:canonical} and obtain consistent results. The spectra derived with and without RSCD masking are compared in Appendix~\ref{app:robustness}. Future work is needed for a detailed assessment of the optimal treatment of the RSCD effect for short MIRI ramps.

In Stage 3, we begin by using the gain reference file to convert the
flux of each pixel from $\mathrm{DN}\,\mathrm{s}^{-1}$ to $\mathrm{e}^-\,\mathrm{s}^{-1}$. Next, we identify outliers caused by cosmic-ray hits by applying sigma clipping to the time series of each pixel. This involves using a running median over seven integrations and setting a threshold of 7$\sigma$. Additionally, we mask neighboring pixels around detected outliers, along with pixels flagged during Stages 1 and 2. To correct for the background and 1/f noise, we subtract the mean flux from the regions outside the trace for each detector column and integration. The background is estimated using columns with pixel numbers 13-29 and 43-59, while ensuring that bad pixels and outliers are excluded. We then extract the spectra using optimal extraction \citep{horne_optimal_1986}, using the median of all integrations as a model for the point-spread function for each spectral channel and an aperture of 9 pixels. During the extraction process, we reject additional outliers and perform further sigma clipping on the resulting light curves, setting the threshold to 5$\sigma$ and 4$\sigma$, respectively. The end products are two time series of each spectral channel, one that includes the standard RSCD step and one that does not.

\subsubsection{Characterizing the settling timescale} \label{sec:time-scale}

\begin{figure}
	\includegraphics[width=0.48\textwidth]{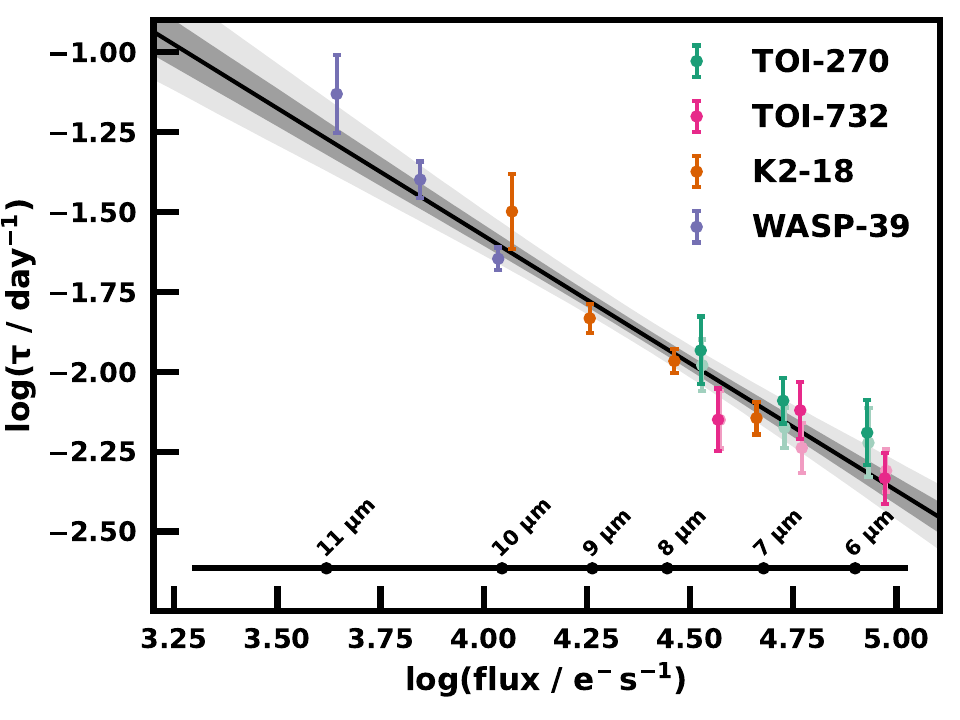}
    \caption{MIRI LRS settling timescale as a function of flux. For the brightest targets, TOI-732 and TOI-270, we also include data points obtained without using the standard RSCD step, shown in lighter colours. The scale at the bottom illustrates the span of flux for the present TOI-270 observation at different wavelengths.}
    \label{fig:exponential_timescale} 
\end{figure}

The MIRI instrument exhibits a characteristic exponential settling ramp at the start of each time-series observation, both when using LRS and for photometry. To correct for this when using MIRI LRS it is typical to fit an exponential ramp together with the astrophysical model for each spectral channel. However, depending on the settling timescale, it can be challenging to precisely fit this exponential ramp if the timescale is sufficiently long or if the signal-to-noise is low (which is often the case for spectroscopically-resolved light curves). As a result, the precision on the transit depth itself can be affected by the precision of the exponentially-decaying baseline, limiting us from achieving optimal signal-to-noise. At the same time, MIRI LRS light curves have been shown to possess time-correlated noise \citep{Madhusudhan2025}, which could be over-fitted by a systematics model that is too flexible \citep[e.g.,][]{Hattori_2022}, potentially leading to inaccuracies in the transmission spectrum. To address these issues, we analyze the exponential settling ramp of a set of MIRI LRS observations to empirically construct a prior on the settling timescale. 

In the case of MIRI F1500W photometry, it was recently reported that the logarithm of the settling timescale is proportional to the magnitude of the target \citep{Connors2025}, suggesting that pixels receiving more flux reach an equilibrium quicker compared to pixels receiving less flux. Therefore, we uniformly analyze MIRI LRS observations of WASP-39~b (DD 2783), K2-18~b (GO 2722), TOI-732~c (GO 3557), in combination with the present TOI-270~d data, to assess if the LRS mode shows a similar trend. These observations cover a wide range of target brightnesses, with 13-100 groups per integration. We reduce the additional data sets following the same method as described in Section~\ref{sec:reduction}. To obtain precise system parameters and limb-darkening coefficients (LDCs), we fit the white light curves of the TOI-270~d, TOI-732~c, and WASP-39~b transits using \texttt{JExoRES}, which utilizes the \texttt{batman} transit model \citep{kreidberg_batman_2015} and nested sampling via \texttt{MultiNest} \citep{Feroz2009}. For K2-18~b, we used the most recent values by \cite{Madhusudhan2025}. For our main target TOI-270~d, we fixed the orbital period $P$ to that of \cite{Mikal-Evans2023} and used the constraints of the scaled semi-major axis $a/R_\star$ and inclination $i$ from \cite{Constantinou2025}, as priors. For TOI-732~c and WASP-39~b, we fixed the orbital periods and used constraints on $a/R_\star$ and $i$ as priors from \cite{Bonfanti2024} and \cite{Carter2024}, respectively. Additionally, we also fit for the secondary eclipse of TOI-732~b which occurred simultaneously as the transit of planet~c. For the eclipse of TOI-732~b, we fit for the eclipse depth, while fixing the orbital parameters and the planet-to-stellar radius-ratio to that of \cite{Bonfanti2024}. We assume circular orbits for all planets. For the exponential trend itself, we used the same parameterization as \cite{Madhusudhan2025}, which combine an exponential and a linear function, via
\begin{equation} \label{eq:trend}
    F(t) = F_\star \, (1 - \alpha (t-t_0) + \beta \, e^{-(t-t_0)/\tau} ) \, F_{\mathrm{transit}}\,,
\end{equation}
where $F_\star$ is the out-of-transit stellar flux, $F_{\mathrm{transit}}$ is the transit model, and $\alpha$, $\beta$, and $\tau$ are trend parameters.

Next, to accurately constrain the settling timescale, we bin the light curves into large $\sim$1~$\mu$m bins, starting at 5.4~$\mu$m. We then fitted these broadband light curves, while fixing the mid-transit times, $a/R_\star$, $i$, and LDCs to the values obtained in the above description. For each target, we used the estimate of the settling timescale $\tau$ from the first 3-4 bins (covering a range of $\sim$5.4-9~$\mu$m). At longer wavelengths, the settling timescale was not well constrained due to low SNR. In Figure~\ref{fig:exponential_timescale}, we show the inferred settling timescales $\tau$ as a function of the average stellar flux within each bin. In this flux range, we find that the settling timescale varies between a few minutes to around 2 hours. This means that it can take $\sim$10 hours for the exponential settling ramp to reach 1\% of the strength at the start of the visit for long wavelengths.

As in the case of F1500W photometry \citep{Connors2025}, we demonstrate that the detector settling timescale strongly depends on the flux when using LRS. We fit a linear relationship to these data using
\begin{equation}
    \log\tau = p_1 \log F_\mathrm{\star} + p_2\,,
\end{equation}
and obtain $p_1=-0.798\pm0.064$ and $p_2 = 1.62 \pm 0.28$. We also include an error inflation parameter, which we add in quadrature to the uncertainties of the timescales. However, we find that this parameter is consistent with zero, with a median value of 0.03. Although we show that this linear relation holds between $\sim$5.4-9~$\mu$m, there may be additional dependencies on other factors, such as the wavelength, PSF shape, or the number of groups, which we do not consider in this work. It is also possible that this relation does not hold for fluxes outside of the tested range; however, the range covered by the selected observations spans most of the flux range of the present TOI-270 observation, as shown in Figure~\ref{fig:exponential_timescale}.

\subsubsection{Light curve fitting} \label{sec:light-curve-fitting}

\begin{table}
\renewcommand{\arraystretch}{1.15} 
\setlength{\tabcolsep}{3pt}
\centering
\begin{tabular}{lcccc}
\hline \hline
Parameter & Prior & Value \\ \hline
P (days) & fixed & 11.38014 \\
T$_0$ (BJD) & $\mathcal{U}$(T$'_0 - 0.02$, T$'_0 + 0.02$) &  $60517.177142_{-0.000050}^{+0.000050}$   \\ 
$i$ (deg) & $\mathcal{N}$(89.756, 0.038) & $89.776_{-0.027}^{+0.029}$   \\ 
$a / R_\star$ & $\mathcal{N}$(42.35, 0.22) & $42.31_{-0.15}^{+0.14}$   \\ 
$R_\mathrm{p} / R_\star$ & $\mathcal{U}$(0.03, 0.07) & $0.05382_{-0.00017}^{+0.00016}$  \\ 
$q_1$ & $\mathcal{U}$(0, 1) & $0.0136_{-0.0055}^{+0.0097}$   \\ 
$q_2$ & $\mathcal{U}$(0, 1)  & $0.47_{-0.26}^{+0.30}$   \\ 
$u_1$ & -- & $0.108_{-0.048}^{+0.040}$   \\ 
$u_2$ & -- & $0.007_{-0.057}^{+0.078}$   \\ \hline
\end{tabular}
\caption{System parameters of TOI-270~d from the \texttt{JExoRES} MIRI white light curve analysis. We fixed the orbital period $P$ to 11.38014 days \citep{Mikal-Evans2023}. For $i$ and $a / R_*$, we used the estimates from \cite{Constantinou2025} as normal priors. We used uniform priors on the mid-transit time T$_0$, corresponding to approximately $\pm$30~minutes from the expected mid-transit time T$'_0$. We give T$_0$ in terms of BJD $-$ 2400000.5 days.
}
\label{tab:wlc_params}
\end{table}

\begin{figure*}
	\includegraphics[width=1\textwidth]{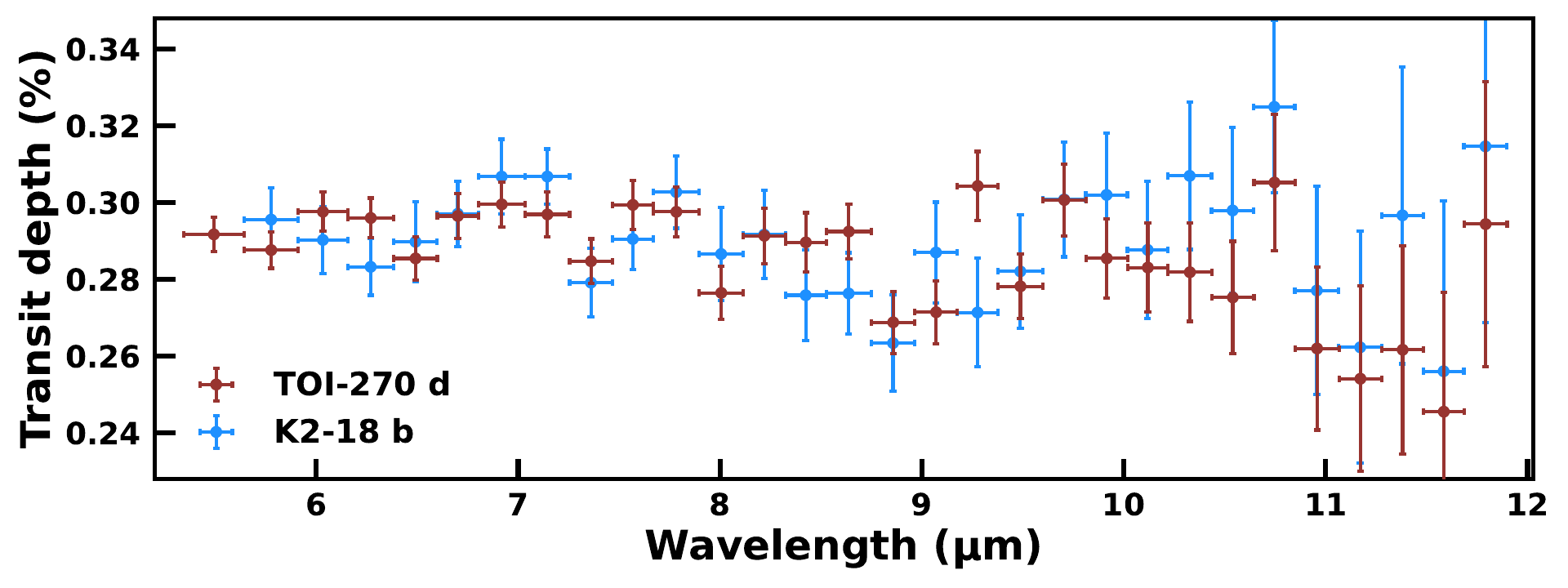}
    \caption{ Comparison between the transmission spectra of TOI-270~d and K2-18~b observed with MIRI LRS. The spectrum for K2-18~b is obtained from \cite{Madhusudhan2025} and is offset by 91~ppm to align with the baseline of the TOI-270~d spectrum. Both spectra were reduced using \texttt{JExoRES}. For TOI-270 d, the RSCD step was not applied.}
    \label{fig:K2_18_comparison} 
\end{figure*}

We fit the white light curve, obtained between 5.4-10~$\mu$m, using the approach outlined above. The parameter estimates from the white light curve fit are presented in Table~\ref{tab:wlc_params}. For the spectroscopic light curve fitting, we bin the data to a bin width of 0.2~$\mu$m or 5 pixels, whichever contains the most pixels \citep[as done in][]{Madhusudhan2025}. We then fix the system parameters and LDCs to the values from the white light curve analysis, while fitting for the planet-to-star radius ratio, the four parameters of Eq. \eqref{eq:trend}, and an error inflation parameter that is added in quadrature. The strategy of fixing the LDCs to the values from the white light curve is adapted from \cite{Madhusudhan2025}. Their work showed that the MIRI spectrum of K2-18~b, which orbits a star similar to TOI-270, is not significantly affected by the treatment of limb darkening. We mask the first 50 integrations of each light curve. We use the relation derived in Section~\ref{sec:time-scale} to construct a set of normally distributed priors on the settling timescale $\tau$. We inflate the prior standard deviation by adding the median error inflation parameter in quadrature (see Section~\ref{sec:time-scale}). After fitting each spectral channel, we re-fit them using the Gaussian process approach in \cite{Madhusudhan2025} to account for time-correlated noise. We note that using the informed prior for the timescale $\tau$ lowers the transit depth uncertainty by an average of 11\% below 10~$\mu$m, and up to $\sim$25\% between 7-9~$\mu$m, as described in Appendix~\ref{app:robustness}. Similar to \cite{Madhusudhan2025}, we do not consider the data at the shortest wavelengths. In fact, we observed persistent outliers in the spectral channel between 5.0-5.3~$\mu$m, which may be due to a large comic ray hit. We show the MIRI transmission spectrum of TOI-270~d in Figure~\ref{fig:K2_18_comparison} and compare it to the K2-18~b spectrum from \cite{Madhusudhan2025}. Note that for the atmospheric retrievals, we remove the data beyond 11~$\mu$m due to low SNR.

\subsection{JexoPipe}
\label{sec:jexopipe}

We perform an additional independent reduction with \texttt{JexoPipe} as a robustness check. \texttt{JexoPipe} is an data reduction framework that utilizes a combination of steps from the JWST Science Cailbration Pipeline and customized steps \citep{Sarkar2024}.  Here we use the MIRI version of \texttt{JexoPipe} as previously applied in \cite{Madhusudhan2025}.

\subsubsection{Data reduction}
\label{jexopipe data reduction}

In \texttt{JexoPipe} stage 1 we take the .uncal files and process these through the following steps from the JWST Science Cailbration pipeline: group scale, data quality initialization, EMI correction, saturation flagging, first frame and last frame correction, reset correction, linearity correction, dark subtraction, ramp fitting and gain scale. The RSCD step was omitted as in the \texttt{JExoRES} reduction.  In stage 2 we combine all segments together and apply the assign WCS (World Coordinate System) and flat field steps. We then apply a custom bad pixel identification step and a custom background subtraction step \citep{Sarkar2024}. In each integration image, we use the same pixel columns as in the \texttt{JExoRES} reduction to sample the background region and, after applying an outlier mask, we obtain the mean of each row. This mean is then subtracted from all pixel values in the corresponding row. We then apply a custom bad pixel correction step, that uses a combination of temporal and spatial interpolation \citep{Sarkar2024}. We next apply an additional outlier correction step, using a rolling median image (of 20
contiguous integrations), to detect 5$\sigma$ outliers and replace them with the pixel value from the rolling median. In stage 3, we apply an extraction aperture centered on the spectral trace, of width 9 pixels. The 1-D stellar spectra are then extracted with an optimal extraction algorithm \citep{horne_optimal_1986}.
 
\subsubsection{Light curve fitting}
\label{jexopipe light_curve_fitting}

 \begin{figure*}
\centering
    \includegraphics[width=0.87\textwidth]{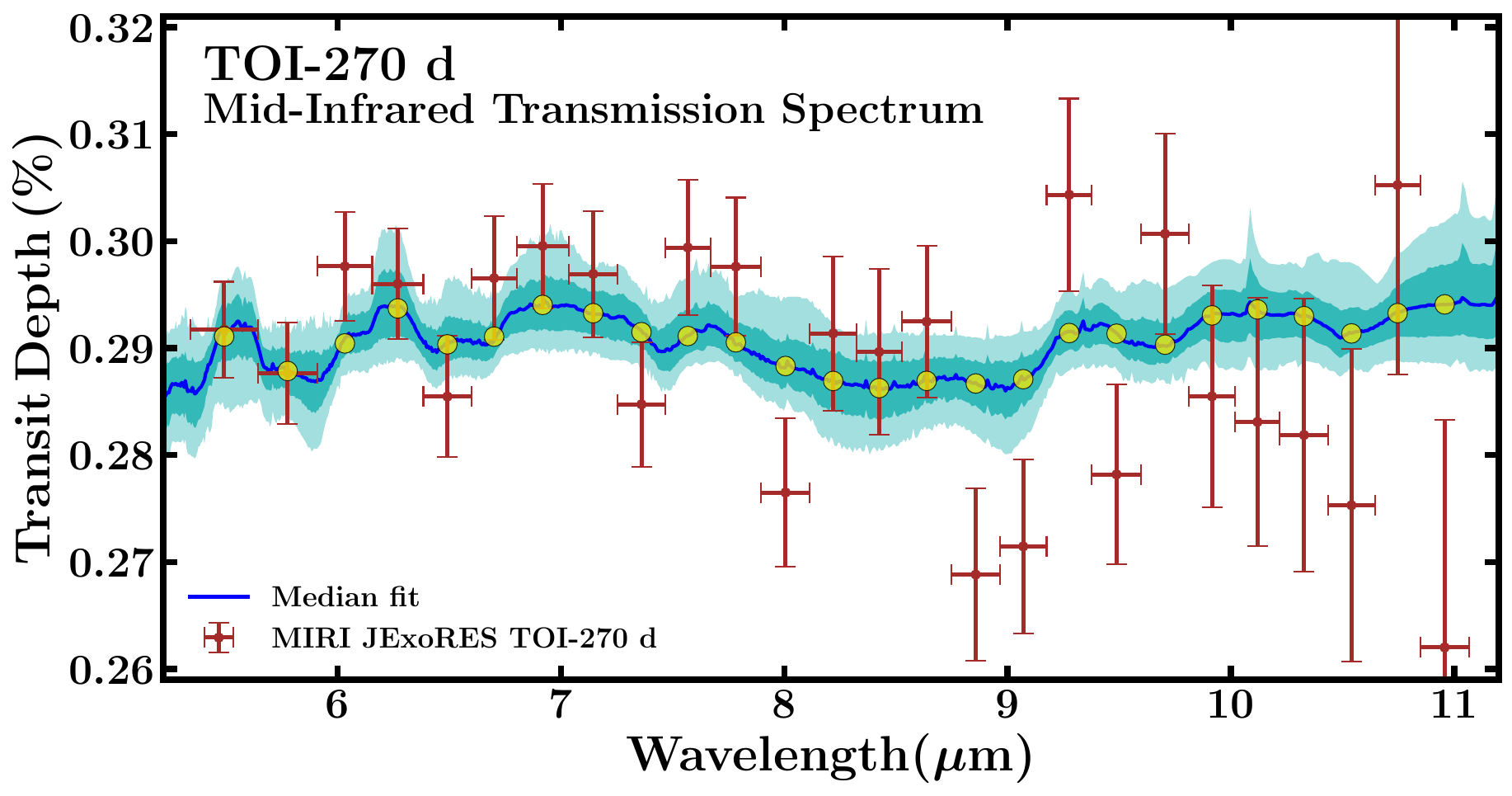}
    \caption{
    The mid-infrared transmission spectrum of TOI-270 d obtained with the JWST MIRI LRS instrument and the \texttt{JExoRES} pipeline. The data points with error bars (in dark red) show the observed spectrum. The horizontal error bars correspond to the spectral bin width. The dark blue curve denotes the median retrieved spectral fit to this MIRI dataset using the canonical retrieval discussed in Section \ref{sec:canonical}, while the two lighter shaded regions denote the 1$\sigma$ and 2$\sigma$ intervals. The yellow points denote the binned median model.
    }
    \label{fig:canonical_plot_JRES} 
\end{figure*}

We extract the white light curve from wavelengths 5-10 \textmu m. We exclude the first 50 integrations which show the steepest trend from detector settling.   We also perform another stage of outlier correction: those integrations $\pm 5\sigma$ from a rolling median on the white light curve are replaced by linear interpolation of the 1-D spectra from adjacent integrations. Initially, the out-of-transit residuals are estimated by fitting a trend function only.  The light curve error bars are inflated so that the average error bar size equals the standard deviation of the residuals. 
We then use \texttt{emcee} \citep{foreman-mackey_emcee_2013} to perform MCMC parameter estimation. We apply a model consisting of a transit model from \texttt{PyLightCurve} \citep{Tsiaras2019} multiplied by the following systematic function:  $a (1+bt +c e^{-t/d})$, where $t$ is the time since the start of the science exposure and $a$, $b$, $c$ and $d$ are coefficients. 
We fit for $R_p/R_s$, $t_0$, $a/R_s$, $i$, two quadratic limb darkening coefficients, and the four trend parameters.  We assume a circular orbit and fix the period to 11.38014 days \citep{Mikal-Evans2023}.
We then proceed to spectral light curve fits.  We first bin the columns to a width of 0.2 $\mu$m or 5 pixels whichever has more pixels, giving a minimum bin width of 0.2 $\mu$m.  We fix $t_0$, $a/R_s$, $i$ and the two quadratic LDCs to the white light values, and fit for $R_p/R_s$ and the four trend parameters. For the time constant $d$ we use Gaussian priors centred on predicted values from Figure \ref{eq:trend}. This gives the final transmission spectrum (shown in Figure \ref{fig:pipeline_comparison}).

\section{Atmospheric Retrievals} \label{sec:retrieval}

We perform atmospheric retrievals using the present MIRI spectrum of TOI-270~d, as described in Section \ref{sec:obs}, as well as recent near-infrared NIRISS SOSS and NIRSpec G395H spectra \citep{Holmberg2024, Constantinou2025}. First, we run a set of `canonical' MIRI retrievals including the 11 species in \cite{Madhusudhan2023b} and three additional species that are important in the MIRI LRS band, based on recent works: DMDS \citep{Schwieterman2024, Tsai2024, Madhusudhan2025}, propyne \citep{Huang2024, welbanks_challenges_2025}, and isoprene \citep{Zhan2021, Schwieterman2024}. Next, we perform an agnostic search of 203 trace species in both the mid- and near-infrared regions, using a similar setup to \cite{pica-ciamarra_systematic_2025} and \cite{Rigby_Madhusudhan_2025}. 

\subsection{Retrieval Setup} 
We employ \texttt{petitRADTRANS} \citep{Molliere2019, Nasedkin2024} for the atmospheric retrievals in this work. We model the terminator atmosphere in one dimension assuming hydrostatic equilibrium and uniform molecular abundances. The atmospheric model consists of 100 layers between 10$^{-6}$ and 10 bar. 
We assume H$_2$ and He (He/H~=~0.17 by number) as the background gas and fit for the mass mixing ratios of additional molecular species.
Since \texttt{petitRADTRANS} (\texttt{pRT}) implements correlated-k opacities natively, we use molecular opacities in the correlated-k mode \citep{Lacis_Oinas_1991} natively available in \texttt{pRT} or obtained from the ExoMol opacity data base \citep{Chubb_Rocchetto_2021, Tennyson_Yurchenko_2024} when available (CH$_4$, CO$_2$, H$_2$O, NH$_3$, CO, CH$_3$F, CS$_2$, CH$_3$Cl, HCN, HF, PH$_3$, SiH$_4$, CH$_2$O, C$_2$H$_2$, C$_2$H$_4$, SO$_2$, SO, SO$_3$, OCS, O$_2$, NS, N$_2$O, NO, NH, PN, H$_2$S, CH$_2$S, CS, SiO), and from HITRAN \citep{Gordon2022} for the rest. For these additional species, we use the cross-sections at $T\approx300$~K and $P \approx 1$~bar when available, or the closest available temperature and pressure, and we convert them to the correlated-k format. The exception is ethane (C$_2$H$_6$) for which we take the HITRAN line list and convert it into a \texttt{pRT} correlated-k table, using Cthulhu \citep{Agrawal_MacDonald_2024} and Exo\_k \citep{Leconte_2021}.
Another difference with most retrieval codes is that the abundance priors for trace species are defined in mass fractions rather than volume mixing ratios (VMR). In order to approximately match the standard bounds in the field for log(VMR) of 10$^{-12}$ to 10$^{-0.3}$ \citep[e.g.,][]{Madhusudhan2023b, Holmberg2024}, we choose mass fraction bounds of 10$^{-11}$ to 1. 
We include collision-induced absorption for H$_2$-H$_2$, H$_2$-He and He-He \citep{borysow1988, orton2007, abel2011, richard_new_2012}. Finally, we also fit the reference pressure between 10$^{-6}$ and 10 bar, which corresponds to R$_p$~=~2.19 R$_\oplus$ \citep{Mikal-Evans2023}. We sample from the posterior probability distribution using \texttt{MultiNest} \citep{Feroz2009}.

\subsection{Canonical Retrievals} 
\label{sec:canonical}

We start with canonical retrievals as described in \cite{Madhusudhan2023b} using the three MIRI spectra from this work: the \texttt{JExoRES} spectra with and without the RSCD step (Section~\ref{sec:time-scale}), and the \texttt{JexoPipe} spectrum (Section~\ref{sec:jexopipe}). We apply an isothermal P-T profile along with a 2-parameter opaque patchy cloud-deck model (i.e., no haze) \citep{macdonald2017, Pinhas2018}. For the isothermal temperature we consider a uniform prior distribution between 0 and 600~K, which is wide enough to cover the planet's zero-albedo equilibrium temperature $T_{eq} = 387$~K \citep{VanEylen2021}. We also consider a non-isothermal P-T profile \citep{Madhusudhan2009}, but find no statistically significant preference over an isothermal P-T profile. We add a Gaussian prior on the planet mass according to the value of $M_p$~=~4.78 $\pm$ 0.43 M$_\oplus$ found in \cite{VanEylen2021}. In each retrieval, we include 14 species: CH$_4$, CO$_2$, H$_2$O, NH$_3$, CO, HCN, N$_2$O, (CH$_3$)$_2$S (DMS), OCS, CS$_2$ CH$_3$Cl, (CH$_3$)$_2$S$_2$ (DMDS), C$_3$H$_4$ (propyne), and C$_5$H$_8$ (isoprene). We summarize the priors and results in Table~\ref{tab:retrieval_priors_canonical} and Figure~\ref{fig:corner-plot_MIRI}. We show our retrieved spectrum on the default MIRI \texttt{JExoRES} dataset (no RSCD step) in Figure~\ref{fig:canonical_plot_JRES}.

\begin{table*}
\centering
\begin{tabular}{l|cc|cccc} \hline  \hline
    Parameter & Prior & Description & \texttt{JexoPipe} & No-RSCD & RSCD \\[0.5mm]
    \hline
    $\mathrm{log}(w_{\mathrm{CH}{_4}})$& $\mathcal{U}$(-11, 0) & Abundance$^*$ of CH$_4$  & $ <-1.91$ & $<-1.73$ & $<-1.69$ \\[0.5mm]
    $\mathrm{log}(w_{\mathrm{CO}_2})$& $\mathcal{U}$(-11, 0) & Abundance of CO$_2$                          & $ <-2.16$ & $<-2.08$ & $<-2.04$ \\[0.5mm]
    $\mathrm{log}(w_{\mathrm{H}_2\mathrm{O}})$& $\mathcal{U}$(-11, 0) & Abundance of H$_2$O                 & $<-1.58$  & $<-1.43$ & $<-1.23$ \\[0.5mm]
    $\mathrm{log}(w_{\mathrm{CO}})$& $\mathcal{U}$(-11, 0) & Abundance of CO                                & $ <-1.81$ & $<-1.97$ & $<-1.96$ \\[0.5mm]
    $\mathrm{log}(w_{\mathrm{NH}_3})$& $\mathcal{U}$(-11, 0) & Abundance of NH$_3$                          & $<-1.54$  & $<-1.39$ & $<-1.41$ \\[0.5mm]
    $\mathrm{log}(w_{\mathrm{HCN}})$& $\mathcal{U}$(-11, 0) & Abundance of HCN                              & $<-1.85$  & $<-2.00$ & $<-1.99$ \\[0.5mm]
    $\mathrm{log}(w_{\mathrm{DMS}})$& $\mathcal{U}$(-11, 0) & Abundance of DMS                              & $<-2.74$  & $<-2.50$ & $<-2.45$ \\[0.5mm]
    $\mathrm{log}(w_{\mathrm{N}_2\mathrm{O}})$& $\mathcal{U}$(-11, 0) & Abundance of N$_2$O                 & $<-2.27$  & $<-2.06$ & $<-2.12$ \\[0.5mm]
    $\mathrm{log}(w_{\mathrm{CS}_2})$& $\mathcal{U}$(-11, 0) & Abundance of CS$_2$                          & $<-2.95$  & $<-2.55$ & $<-2.27$ \\[0.5mm]
    $\mathrm{log}(w_{\mathrm{CH}_3\mathrm{Cl}})$& $\mathcal{U}$(-11, 0) & Abundance of CH$_3$Cl             & $<-2.31$  & $<-2.37$ & $<-2.58$ \\[0.5mm]
    $\mathrm{log}(w_{\mathrm{OCS}})$& $\mathcal{U}$(-11, 0) & Abundance of OCS                              & $<-2.10$  & $<-2.17$ & $<-1.96$ \\[0.5mm]
    $\mathrm{log}(w_{\mathrm{DMDS}})$& $\mathcal{U}$(-11, 0) & Abundance of DMDS                            & $<-2.56$  & $<-2.52$ & $<-2.54$ \\[0.5mm]
    $\mathrm{log}(w_{\mathrm{Propyne}})$& $\mathcal{U}$(-11, 0) & Abundance of Propyne                      & $<-3.72$  & $<-2.70$ & $<-2.26$ \\[0.5mm]
    $\mathrm{log}(w_{\mathrm{Isoprene}})$& $\mathcal{U}$(-11, 0) & Abundance of Isoprene                    & $-3.28_{-5.90}^{+1.48}$ & $-3.55_{-6.43}^{+1.80}$ & $-4.20_{-6.80}^{+2.44}$           \\[0.5mm]
    $\mathrm{log}(P_\mathrm{ref}/\mathrm{bar})$ & $\mathcal{U}$(-6, 1) & Reference pressure at R$_\mathrm{p, ref}$   & $-3.26_{-1.08}^{+1.65}$ & $-3.05_{-1.15}^{+1.76}$ & $-2.47_{-1.45}^{+1.73}$  \\[0.5mm]
    $T_\mathrm{iso} / \mathrm{K}$ & $\mathcal{U}$(0, 600) & Isothermal temperature                                              & $221_{-94}^{+120}$     & $210_{-95}^{+143}$      & $197_{-95}^{+150}$        \\[0.5mm]
    $\mathrm{log}(P_\mathrm{c}/\mathrm{bar})$ & $\mathcal{U}$(-6, 1) & Cloud top pressure                            & $-2.62_{-1.88}^{+1.99}$ & $-2.52_{-1.91}^{+2.18}$ & $-2.49_{-2.07}^{+1.93}$  \\[0.5mm]
    $\phi$ & $\mathcal{U}$(0, 1) & Cloud/haze coverage fraction                                                      & $0.47_{-0.29}^{+0.29}$  & $0.48_{-0.29}^{+0.30}$  & $0.51_{-0.29}^{+0.28}$   \\[0.5mm]
    \hline
    ln Z & & Log evidence & 195.10 & 200.76 & 194.52  \\ \hline
\end{tabular}
\caption{Canonical retrieval prior distributions and results for each case. The RSCD and no-RSCD (default) cases correspond to the \texttt{JExoRES} MIRI spectra.
$^*$ The priors are in mass fractions, but the posteriors were converted in volume mixing ratios for easier comparison with previous works.
We provide the median values with 1$\sigma$ error bars for isoprene which has a peaked posterior, and 95\% upper limits for all other species.
}
\label{tab:retrieval_priors_canonical}
\end{table*}

Using the above setup, we find nominal contributions from isoprene in all three MIRI reductions, and no additional constraints on any other molecules in the model. With \texttt{JExoRES}, we find consistent volume mixing ratio for isoprene between the two reductions, obtaining log(X$_{\mathrm{Isoprene}}$) of $-4.20_{-6.80}^{+2.44}$ (ln~$B = 0.46$) and $-3.55_{-6.43}^{+1.80}$ (ln~$B = 1.01$) with and without the RSCD step, respectively, while with \texttt{JexoPipe} we obtain $-3.28_{-5.90}^{+1.48}$ (ln~$B = 0.79$). These Bayes factors indicate that the evidence for isoprene is at most weak when using the canonical model. However, when considering a reduced model containing only CH$_4$, CO$_2$, and H$_2$O, the inclusion of isoprene yields higher Bayes factors (ln~$B = 1.34–2.64$), depending on the data reduction and model assumptions, as discussed in Section~\ref{sec:exploration}. We only find upper limits of CH$_4$, CO$_2$, H$_2$O, or CS$_2$ using this setup, for which previous works have reported evidence using near-infrared data \citep{Holmberg2024, Benneke2024, Felix2025, Constantinou2025}. The lack of constraints for CH$_4$ and CO$_2$ are not surprising given that their cross-sections are relatively smaller in the MIRI LRS range compared to other species, such as Isoprene, as seen in Figure \ref{fig:contribution_plot_MIRI}. This is consistent with recent inferences using the MIRI spectrum of K2-18~b \citep{Madhusudhan2025}. While the canonical model does not yield significant evidence for H$_2$O, model comparisons considering only CH$_4$, CO$_2$ and H$_2$O contributions provides weak evidence for H$_2$O (ln~$B = 1.68$), as shown in  Section~\ref{sec:exploration}. We also place a 95\% upper limit on log(X$_{CS_2}$) between $-2.27$ and $-2.95$, which is consistent with previous studies. For the temperature, we obtain consistent estimates of around 200-220$_{-100}^{+150}$~K, consistent with the NIR \citep[e.g.,][]{Constantinou2025}. Finally, we do not find any significant constraints on clouds using the MIRI data alone.

\begin{table*}
\renewcommand{\arraystretch}{1.05} 
\setlength{\tabcolsep}{13pt}
\centering
    \begin{tabular}{l|ccc} \hline \hline
         Molecule & MIRI &  MIRI & NIR \\
                & \texttt{JExoRES} & \texttt{JexoPipe} &  \\
         \hline
         Benzonitrile (C$_7$H$_5$N)        & $-4.31_{-1.06}^{+1.25}$ (3.43, 4.32) & $-3.97_{-1.08}^{+1.33}$ (5.25, 6.01) & $<-5.01$ (0.08, 0.13) \\
         Methylacrylonitrile (C$_4$H$_5$N) & $-3.37_{-1.56}^{+1.14}$ (2.48, 3.06) & $-3.87_{-1.32}^{+1.14}$ (3.22, 3.87) & $<-4.96$ (0.36, -0.31) \\ 
         Isobutene (C$_4$H$_8$)            & $-3.91_{-1.63}^{+1.40}$ (2.32, 3.08) & $-4.23_{-1.32}^{+1.29}$ (4.27, 5.02) & $<-4.84$ (0.25, 2.37) \\ 
         Styrene (C$_8$H$_8$)              & $-3.67_{-1.74}^{+1.21}$ (1.71, 2.78) & $-3.10_{-1.70}^{+0.81}$ (1.76, 2.62)& $<-5.56$ (0.21, 0.96) \\
         Cumene (C$_9$H$_{12}$)            & $-4.66_{-1.73}^{+1.60}$ (1.40, 2.19) & $-4.55_{-1.37}^{+1.33}$ (2.90, 3.07)& $<-5.11$ (-0.01, 0.46) \\
         Isoprene (C$_5$H$_8$)             & $-3.52_{-2.41}^{+1.31}$ (1.34, 2.19) & $-3.45_{-1.98}^{+1.17}$ (2.04, 2.64)& $<-4.74$ (0.40, 1.23) \\ \hline
    \end{tabular}
    \vspace{2mm}
    \caption{Volume mixing ratio posteriors for the molecules that reach ln~$B$ $\geq$ 2 with both MIRI spectra, as detailed in Section~\ref{sec:exploration}. In brackets, we show the ln B value for the default (cloudy) retrieval followed by that for a cloud-free case. We find no significant evidence for the presence of clouds in the MIRI data but consider the cloudy case as a conservative measure which provides lower Bayesian evidence due to the additional parameters which affect the spectral baseline.
    When a posterior is unconstrained, we give the 2$\sigma$ upper bound. The full table of model preferences for the cloudy case is given in the Appendix (Table~\ref{tab:all-molecules}). We see that none of these species reaches the threshold of ln~$B$ $\geq$ 2 in the near-infrared spectrum when including clouds.}
    \label{tab:exploration-MIRI}
\end{table*}

In addition to the 14 molecules included in the canonical model, we also perform retrievals with a larger set of molecules. For this, we include 7 additional molecules: H$_2$S, SO$_2$, C$_2$H$_2$, C$_2$H$_4$, C$_2$H$_6$, CH$_4$S, PH$_3$, similar to the maximal model in \cite{Madhusudhan2025}. From this maximal retrieval using the \texttt{JExoRES} spectrum, we found that out of the 21 molecules, only Isoprene displayed a well-defined peak in the posterior distribution, which aligns with the findings of the canonical model. We now proceed to conduct a more extensive exploration in search of other candidates beyond these molecules, which may be degenerate with Isoprene or with additional contributions to the observed spectrum.

\subsection{MIR Exploration of Molecular Species} 
\label{sec:exploration}

We conduct a comprehensive search for trace species in the near- and mid-infrared data. Starting with MIRI, we use both the default \texttt{JExoRES} spectrum without the RSCD step (which we will simply refer to as the \texttt{JExoRES} spectrum in this section) and the \texttt{JexoPipe} spectrum, as described in Section~\ref{sec:obs}. We include CH$_4$, CO$_2$ and H$_2$O, which have been detected in the near-infrared \citep{Holmberg2024, Benneke2024}, and perform an agnostic search for additional species following the approach of \cite{Madhusudhan2025} and \cite{pica-ciamarra_systematic_2025}. Although the detection significance for H$_2$O in TOI-270~d in previous works was only $\simeq 2\sigma$, we include it because it is the most dominant O-bearing molecule expected in H$_2$-rich atmospheres, and the data allow for high abundances, unlike in K2-18~b which showed a stringent upper limit.
We search for one additional molecule at a time, for a total of 3+1 molecules per retrieval. This is referred to as the 3+X retrieval setup where X is any of the chemical species that we search for, following the nomenclature of \cite{pica-ciamarra_systematic_2025} and \cite{Rigby_Madhusudhan_2025}. 

In total, we search for 203 molecules, including 90 hydrocarbons, 36 sulfur-bearing species, 64 nitrogen-bearing species, 7 halogen-bearing species and 6 others (CO, CH$_2$O, O$_2$, SiO, SiH$_4$, PH$_3$). 
For this, we use 500 live points and the same isothermal P-T profile and cloud model as in the canonical case, for a total of 8 free parameters (7 for the baseline retrieval). Moreover, these parameters have the same priors as in the canonical retrievals (see Table~\ref{tab:retrieval_priors_canonical}). While we don't find significant evidence for the presence of clouds in the MIRI data we include them in the model to still allow for the possibility and to be conservative in the estimation of the detection significances. The inclusion of the cloud parameters, which affects the baseline of the spectrum, slightly decreases the Bayesian evidence for a given molecule. 

\begin{figure*}
\centering
	\includegraphics[width=0.87\textwidth]{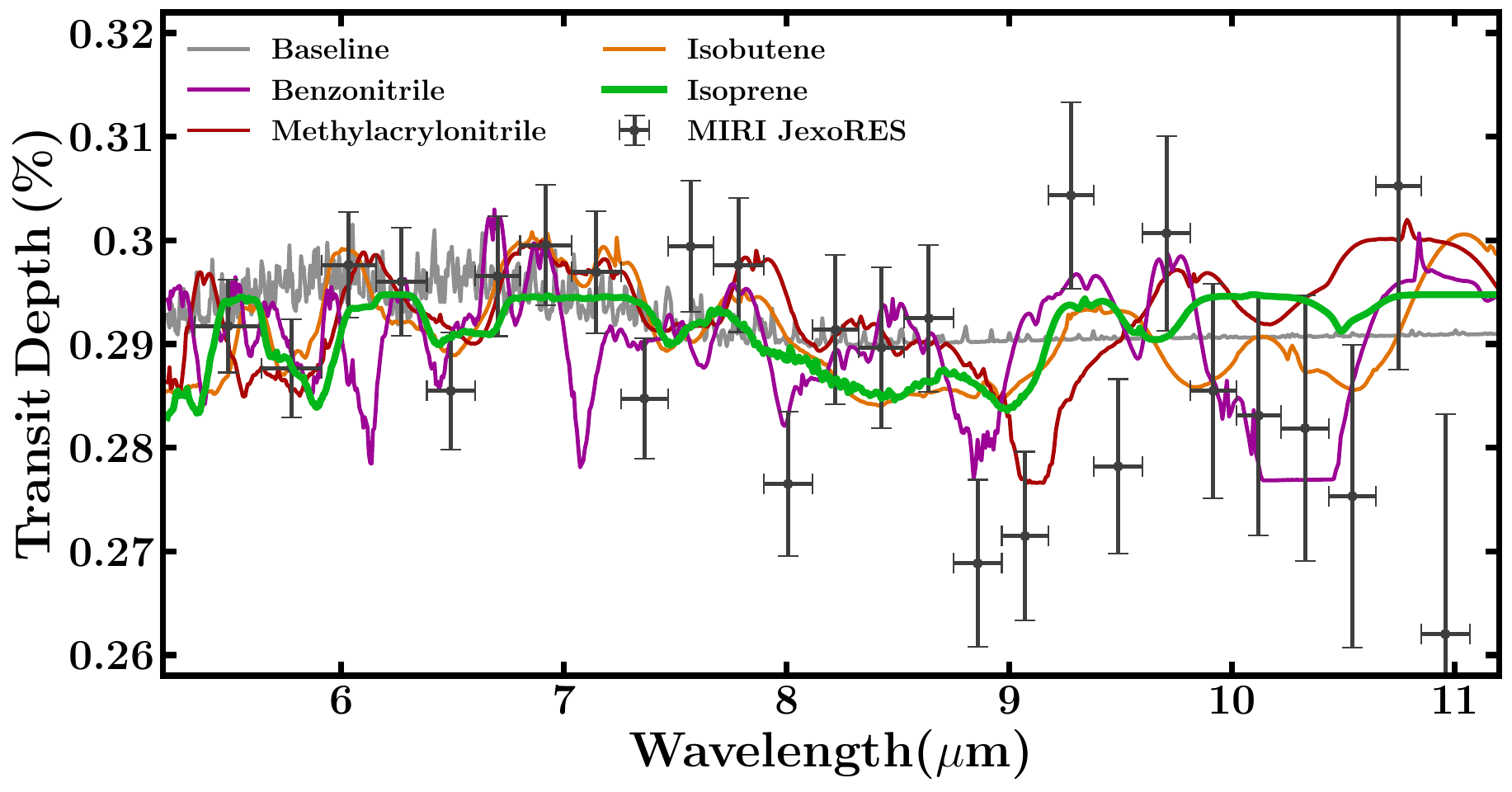}
     \caption{
     Retrieved spectra with the JWST MIRI LRS \texttt{JExoRES} data. The solid curves show median retrieved spectra obtained from the \texttt{pRT} retrievals conducted for the 3 species that reach moderate evidence, as described in Section~\ref{sec:exploration}. To these 3 species, we add Isoprene that reaches weak evidence (ln $B$ $\geq$ 1) and has been proposed as a potential biomarker for Earth-like and H$_2$-rich atmospheres \citep{Zhan2021}. Also shown are the MIRI observations from the \texttt{JExoRES} reduction.
     }
     \label{fig:fit_plot_MIRI} 
\end{figure*}

In Table~\ref{tab:exploration-MIRI}, we present all species that achieved a logarithmic Bayes factor (ln $B$) $\geq$ 2 compared to a base model with only CH$_4$, CO$_2$ and H$_2$O included, with or without considering clouds. This corresponds to a preference of around 2.5$\sigma$ using the inverse conversion of \cite{Sellke_Bayarri_2001}. However, recent studies have questioned the reliability of this conversion in Bayesian inference, e.g. \citet{Thorngren_Sing_2025}. We, therefore, report the $\ln B$ values throughout most of this work. Figure~\ref{fig:fit_plot_MIRI} shows the median model fits for several of these species. We note that this does not mean that these species are detected, given the degeneracies between the spectral features. Rather, these represent candidate species that could explain the data above our evidence threshold.

We find that a few other species provide similar or better fits to the data compared to isoprene that was evident based on the canonical model alone. Three species achieve ln $B$ $\geq$ 2 using both MIRI reductions when considering clouds and six species without clouds. These are all complex molecules (with 10 atoms or more), and none except isoprene \citep{Zhan2021} have significant sources known in planetary atmospheres. One of these complex molecules is isobutene, for which we retrieve a log mixing ratio between $-3.91_{-1.63}^{+1.40}$ and $-4.23_{-1.32}^{+1.29}$. Isobutene could potentially form as a photochemical product of methane, but is unlikely to reach the inferred high abundance without other lower-order hydrocarbons being also being present at even higher abundances \citep{Moses_Fouchet_2005, Nixon_2024}. It is interesting to note that isobutene was also one of the two complex molecules that showed moderate evidence across all datasets of TOI-732~c \citep{Rigby_Madhusudhan_2025}. Next, we have two molecules that lack known natural production mechanisms in the solar system: benzonitrile and methylacrylonitrile. These molecules are produced industrially on Earth. Beyond benzonitrile, methylacrylonitrile and isobutene, three more molecules reach $\ln B$ above 2 for the cloud-free case: styrene, cumene and isoprene. Again, both styrene and cumene have no significant sources known in planetary atmospheres. Finally, isoprene is a proposed biomarker for Earth-like and H$_2$-rich atmospheres \citep{Zhan2021, Schwieterman2024}, for which we retrieve a log mixing ratio between $-3.08_{-2.35}^{+1.26}$ and $-4.94_{-0.77}^{+0.43}$, consistent with constraints from the canonical retrieval. This high abundance could potentially accumulate due to a sufficiently high flux of isoprene, resulting in photochemical self-shielding \citep{Zhan2021}, similar to what has been proposed for DMS and other molecules \citep{Tsai2024}.

None of these six molecules is robustly identified in the near-infrared with a $\ln B$ above 2. However, a few species with somewhat lower $\ln B$ in the MIRI data do appear with higher evidence in the NIR data, such as DMDS and ethyl cyanide, as discussed below. Apart from these species, we also find a ln~$B = 1.68$ preference for H$_2$O using the MIRI \texttt{JExoRES} spectrum, obtained by comparing our baseline model with and without H$_2$O. Our search also includes DMS and DMDS which have been inferred previously for the candidate Hycean world K2-18 b \citep{Madhusudhan2025}. For DMDS we obtain $\ln B$ between 0.62-1.02 whereas for DMS it is 0.39-0.43. We present the relative evidence for all 203 species in Tables~\ref{tab:all-molecules} and~\ref{tab:all-molecules-NIR}.

One caveat with this mid-infrared exploration is the spectra do not significantly constrain the abundances of the base species CH$_4$, CO$_2$ or H$_2$O that are found in the near-infrared. Knowing this, we can use the near-infrared constraints to put narrower priors on these base species for the mid-infrared analysis. We consider lower prior bounds on the volume mixing ratio of $10^{-3}$ for CH$_4$ and $10^{-5}$ for CO$_2$ and H$_2$O to encapsulate the near-infrared constraints \citep[e.g.,][]{Constantinou2025}. We do this for the 9 species with $\ln B \geq 1$ across both MIRI spectra in the initial exploration, and also show the results in Table~\ref{tab:all-molecules}. We see that the model preferences are all slightly reduced, but still very consistent with the initial analysis. Only Isobutene drops below the $\ln B$ threshold of 2, at a value of 1.79 for the \texttt{JExoRES} spectrum. Using these priors, we obtain a MMW of $3.37_{-0.75}^{+1.61}$~amu for the \texttt{JExoRES} spectrum when including isoprene, consistent with the near-infrared constraint of $4.84_{-1.08}^{+1.10}$~amu from \cite{Constantinou2025} within the uncertainties. We further test the sensitivity of our results to the CS$_2$ prior by additionally imposing a lower bound of $10^{-7}$ on its volume mixing ratio for the same set of nine species. This has only a minor impact on the inferred model preferences; the $\ln B$ values for \texttt{JExoRES} decrease by 0.04 on average, while for \texttt{JexoPipe} they increase by 0.4.

\subsection{NIR Exploration of Molecular Species} 

For the near-infrared search, we use both the NIRISS and the NIRSpec spectra \citep{Holmberg2024, Constantinou2025}. We use the same setup as for the MIRI search, only we introduce an offset between NIRISS and NIRSpec, giving us 9 free parameters (8 for the baseline retrieval). We find 11 species with $\ln B \geq 2$, with the highest value for CS$_2$ ($\ln B$~=~5.87), which was also inferred in multiple previous studies \citep{Holmberg2024, Benneke2024, Felix2025, Constantinou2025}, followed by CH$_3$F ($\ln B$~=~4.44) which was one of the potential species noted in \cite{Felix2025}. We also find $\ln B \geq$ 3 for ethyl cyanide and thionyl fluoride that were not previously searched for in TOI-270~d. We show the spectral contributions of these four species in Figure~\ref{fig:contribution_plot_NIR}.

We also find $\ln B \geq$ 1 for ethane (C$_2$H$_6$), which is consistent with the findings of \cite{Constantinou2025}, and for thioformaldehyde (H$_2$CS), as noted by \cite{Felix2025}. Additionally, we observe that dimethyl disulfide (DMDS) and ethyl cyanide are favored at $\ln B \geq$ 2 in the NIR spectrum and at $\ln B \geq$ 1 in the MIRI \texttt{JexoPipe} spectrum, although they only reach $\ln B$ values of 0.62 and 0.96 in the \texttt{JExoRES} spectrum, respectively. None of the molecules achieve $\ln B \geq$ 2 in both the NIR and MIR spectra. Therefore, it is possible that the excess absorption may be due to other species not considered here or that it may require a combination of different molecules to adequately explain all the data.

Finally, because of the high model preference for CS$_2$, we re-analyze the species with $\ln B \geq 2$, as well as ethane, by adding CS$_2$ to the baseline model, hence running ``4+X" (CH$_4$ + CO$_2$ + H$_2$O + CS$_2$ + X) retrievals. We also show these results in Table~\ref{tab:all-molecules-NIR}, next to the 3+X results. Only CH$_3$F ($\ln B$~=~2.99) and ethyl cyanide ($\ln B$~=~2.53) remain at $\ln B \geq 2$, while ethane is the only species for which the preference increased when including CS$_2$, reaching $\ln B = 1.83$.

Our extensive exploration of possible molecules marks the first step in evaluating the evidence for additional, potentially complex species in the atmosphere of TOI-270~d in the mid-infrared. Factors such as integrating both near- and mid-infrared data, or considering multiple molecules simultaneously might influence which molecules are favored. Future theoretical studies and follow-up observations are necessary to reliably determine the presence and abundance of molecules beyond CH$_4$, CO$_2$ and H$_2$O in the atmosphere of TOI-270~d.

\subsection{Is the MIRI Spectrum Featureless?} \label{sec:spectral_features}

We have demonstrated that the MIRI spectrum of TOI-270~d can be explained by several different molecules, which provide a better fit to the data compared to CH$_4$, CO$_2$, and H$_2$O alone. In line with recent works that interpret the MIRI spectrum of K2-18~b \citep{Madhusudhan2025, Taylor2025, welbanks_challenges_2025}, we go on to perform a series of tests to establish whether the spectral features are significant. The first test involves checking whether we can reject a featureless spectrum from the $\chi^2$ value of the residuals between the data and a constant transit depth model, i.e. a flat-line model, which we fit to the data. We note that the flat-line model is unphysical in this context, as described below. Nevertheless, we conduct this test for an initial assessment independent of an atmospheric model. For the \texttt{JExoRES} spectra, we find $\chi^2=47.0$ and $\chi^2= 40.3$ for the cases with and without the RSCD step, respectively, for 25 degrees of freedom using the data shown in Fig.~\ref{fig:canonical_plot_JRES}. These values correspond to p-values between 0.0050 and 0.027, indicating that the data reject the null hypothesis of a flat line at a p-value threshold of 0.05. Although not applicable here, we note that it is possible to mistake spectrally coherent signals for noise given that this analysis implicitly assumes that the spectral channels are independent. Consequently, a low $\chi^2$ value does not necessarily rule out the possibility of spectral features, in contrast to what is suggested in \cite{Taylor2025} and \cite{welbanks_challenges_2025}. This limitation can be addressed by performing atmospheric retrievals or employing methods such as cross-correlation, which can match the structure in the data using physically motivated models.

Another approach is to compare the Bayesian evidence of an atmospheric model with that of a one-parameter model with a wavelength-independent transit depth, as pursued in \cite{Madhusudhan2025}. However, we note that this comparison is not justified given the robust detection of spectral features in the near-infrared, making the flat-line model incorrect a priori. Using the NIR spectrum, a flat-line model is ruled out conclusively at $\ln B = 63$. Moreover, since two independent NIR observations both exhibit clear spectral features, it is unlikely that the spectrum would be completely featureless in the present MIRI epoch due to the variability of, for example, high-altitude clouds. For completeness, we nevertheless perform this test on the MIRI data. Using this simplistic ``flat-line'' model, we obtain $\ln Z$ values (where $Z$ is the Bayesian evidence) between 191.4-192.5 and 197.1-198.3 for the \texttt{JExoRES} spectra with and without the RSCD step. The spread in $\ln Z$ reflects the range of the assumed prior on the constant transit depth, which we assume to be uniformly distributed around the white light curve transit depth $\delta = 2897$~ppm. We consider priors with bounds of $\delta\pm$600~ppm and $\delta\pm$2000~ppm as two end-member scenarios \citep{Madhusudhan2025}. The canonical model is favored over the flat-line model with $\ln B$ values of 2.0-3.1 and 2.5-3.7, respectively, for \texttt{JExoRES} with and without the RSCD step. However, given that the flat-line model is physically implausible and the inferred Bayes factors depend strongly on the priors, this comparison does not provide a reliable measure of the presence of spectral features.

Instead, a more appropriate test is to compare physically plausible atmospheric models with and without molecular absorption. We therefore run a retrieval assuming a pure H$_2$/He atmosphere, adopting identical prior ranges to the canonical model for all other parameters, including clouds and the P-T profile.  We note that this featureless atmospheric model contains additional parameters compared to the above flat-line model and is therefore expected to yield a somewhat lower Bayesian evidence, even if both produce similarly flat spectra. In this context, “featureless” refers to the absence of molecular line absorption; the continuum level is instead set by an opaque cloud deck, such that the resulting spectrum is effectively flat. For the \texttt{JExoRES} spectra, with and without the RSCD step, we obtain $\ln B$ values of 2.8 and 3.4, respectively, when comparing the canonical model to this featureless atmospheric model. Finally, several molecules provide a better fit than those included in the canonical model, as described in Section~\ref{sec:exploration}. As a result, when comparing the highest-evidence model (benzonitrile, see Table~\ref{tab:all-molecules}) with the featureless model, we obtain $\ln B$ = 5.3 for \texttt{JExoRES} without the RSCD step. Taking all of this together, we conclude that the present MIRI data of TOI-270~d favor an atmospheric model with spectral features over one without at $\ln B$ values of $2.8$ to $5.3$, comparable to a similar inference for K2-18~b \citep{Madhusudhan2025}.

\subsection{Comparison with K2-18~b}

Next, we compare the present spectrum of TOI-270~d to the MIRI spectrum of K2-18~b \citep{Madhusudhan2025}, given similarities in the near-infrared spectra of these planets \citep{Holmberg2024}. Figure~\ref{fig:K2_18_comparison} illustrates the similarity between the two MIRI spectra, with most data points being consistent after applying a small offset. To quantify this, we compute the reduced $\chi^2$ between the two spectra (after fitting for an offset) and compare this to the distribution of reduced $\chi^2$ values assuming that both spectra are pure noise. We find a p-value of 0.09, indicating that the two spectra are consistent within the current measurement precision. The fact that both spectra are similar and that they appear to be inconsistent with being featureless \citep[$\sim$3$\sigma$; ][]{Madhusudhan2025} suggest that the spectra are not just random noise, in contrast to \cite{Taylor2025}, motivating further MIRI observations of both TOI-270~d and K2-18~b. 

The central question thus becomes: are these observed features of atmospheric origin, or can they be attributed to potential systematics \citep[e.g.,][]{Stevenson2025}? If it is systematics, we would expect to find other MIRI LRS exoplanet spectra with features similar to those of TOI-270~d and K2-18~b. So far, this appears not to be the case. For example, the hot Jupiter WASP-39~b was reported to have SO$_2$ features between 7-10~$\mu$m and a drop in the transit depth at $>$10~$\mu$m \citep{Powell2024}, attributed to contamination from a debris disk \citep{Flagg2024}, none of which are observed in the present data. This may be explained if the potential systematic errors are more pronounced for targets with smaller transit depths than WASP-39~b, such as TOI-270~d. If this is the case, we expect that the MIRI LRS transmission spectrum of the super-Earth L~168-9~b, with a transit depth much smaller than TOI-270~d, to show clear signs of these systematics, with a spectrum similar to TOI-270~d and K2-18~b. Again, this does not appear to be the case \citep{Bouwman2023, Alam2025}. Another possibility is that stellar contamination is affecting the spectra; however, this is unlikely given that MIR wavelengths are less affected by stellar inhomogeneities than NIR, where no evidence for stellar contamination was found for either TOI-270~d or K2-18~b \citep{Madhusudhan2023, Holmberg2024}. Furthermore, if stellar contamination were a key factor, the spectrum of L~168-9~b could also have been affected, given that this planet also orbits an M-dwarf. Therefore, we conclude that systematic errors are unlikely to significantly affect the MIRI spectra of TOI-270~d and K2-18~b. This leaves genuine atmospheric features as a plausible explanation. However, more observations are needed to assess this more rigorously, given the low SNR data.

Similarly to K2-18~b, we find evidence for additional absorption features in the spectrum of TOI-270~d other than the prominent species (CH$_4$, CO$_2$, and H$_2$O). Notably, some molecules potentially identified in the spectra of K2-18~b and TOI-732 c also appear to be candidates for explaining the MIRI spectrum of TOI-270~d, e.g., methylacrylonitrile and isobutene \citep{pica-ciamarra_systematic_2025, Rigby_Madhusudhan_2025}. Overall, the similarities between the MIRI transmission spectra of TOI-270~d and K2-18~b provide compelling evidence that we are indeed seeing actual mid-infrared spectral features rather than noise, albeit at low SNR. Our work highlights that K2-18~b and TOI-270~d stand out as excellent targets for follow-up observations using mid-infrared transmission spectroscopy with JWST.

\section{Summary and Discussion} \label{sec:discussion}

In this work, we present the first mid-infrared transmission spectrum of TOI-270~d, observed with JWST MIRI LRS. We conducted the data reduction using two independent pipelines and performed several robustness tests. We also established a relationship between the detector settling time and the flux by analyzing four MIRI LRS observations of different targets. This latter analysis improved the precision of the transmission spectrum by an average of 11\% below 10~$\mu$m, demonstrating a strategy that can be applied to other MIRI LRS data sets. We performed a series of atmospheric retrievals, exploring 203 potential species with both the current MIRI spectrum and previously published near-infrared data by \cite{Holmberg2024} and \cite{Constantinou2025}. Similar to the recent analysis of the mid-infrared spectrum of K2-18~b \citep{Madhusudhan2025, welbanks_challenges_2025, pica-ciamarra_systematic_2025}, we find that the MIRI spectrum of TOI-270~d can be explained by any one of 3-6 complex molecules beyond CH$_4$, CO$_2$, and H$_2$O. Notably, some of the best-fitting species identified include benzonitrile, methylacrylonitrile, isobutene, and potentially isoprene. Follow-up observations of TOI-270~d are necessary to definitively determine the presence and abundance of additional species in its atmosphere. 

We find that the present MIRI spectrum of TOI-270~d contains notable evidence for excess absorption due to complex molecules. Considering atmospheric models with and without molecular line absorption, we find evidence for spectral features at $\ln B$ of 2.8-5.3. By comparing the mid-infrared spectrum of TOI-270~d to that of K2-18~b from \cite{Madhusudhan2025}, we also find that the two spectra are consistent after applying a small offset. Together, we conclude that the current MIRI spectrum provides significant evidence of genuine atmospheric signatures which are unlikely to be explained by systematics.

Near-infrared observations of TOI-270~d have revealed an H$_2$-rich atmosphere containing CH$_4$, CO$_2$, and potentially H$_2$O and CS$_2$ \citep{Holmberg2024, Benneke2024}. There is also growing evidence for additional absorbers \citep{Felix2025, Constantinou2025}, which our work adds to. In the present work, we identify several new molecules that may explain the planet's transmission spectrum in the mid-infrared. Of the full range of molecules, we find 3-6 molecules that reach at least $\ln B \geq$ 2 using the MIRI data alone. When combined with NIR data, we find that none meets the same criteria for both NIR and MIR. The molecules that came closest to meeting the criteria are DMDS and ethyl cyanide, which reach $\ln B \geq$ 2 in the NIR and $\ln B \sim$ 1 in the MIR. We also note that the excess absorption may be explainable by opacity sources not considered in this study or may require a combination of different molecules to accurately explain all of the data.

Future work is required to assess the viability of the inferred candidate species. With the exception of isoprene, which has been predicted to be a biomarker \citep{Zhan2021, Schwieterman2024}, these species represent complex hydrocarbons that have not been predicted in exoplanet atmospheres and have no strong abiotic sources on Earth. More work is needed to increase the SNR of the observations in order to robustly identify bonafide molecular absorbers and to establish their plausible sources in TOI-270~d. Overall, we find evidence for excess absorption in the transmission spectrum of TOI-270~d that does not originate from typical sources predicted by atmospheric chemistry, motivating future theoretical , experimental, and observational studies.

\vspace{3mm} {\it Acknowledgments:} This work is based on observations made with the NASA/ESA/CSA James Webb Space Telescope as part of Cycle 2 GO Program 3557 (PI: N. Madhusudhan).  We thank NASA, ESA, CSA, STScI, everyone whose efforts have contributed to the JWST. This work is supported by a research grant to N.M. from the UK Research and Innovation (UKRI) Frontier Grant (EP/X025179/1). 
N.M. and F.E.R. acknowledge support from UKRI STFC toward the doctoral studies of F.E.R (UKRI grant 2605554).
N.M. and M.B. acknowledge support from the UKRI toward the doctoral studies of M.B (Frontier grant EP/X025179/1).
J.M. acknowledges support from NASA Exoplanets Research Program 80NSSC23K0281 and JWST-GO-03557, which was provided by NASA through a grant from the Space Telescope Science Institute, which is operated by the Association of Universities for Research in Astronomy, Inc., under NASA contract NAS 5-03127. 

\vspace{3mm} {\it Author Contributions:} N.M. conceived, planned and led the project. N.M. led the JWST proposal with contributions from S.S., F.E.R and J.M. M.H. and S.S. conducted the data reduction and analyses. M.B. conducted the atmospheric retrievals with inputs from N.M and M.H. M.H. led the writing of the manuscript with contributions from N.M, M.B., S.S. and F.R. All authors provided comments on the manuscript. 

\vspace{3mm} {\it Data Availability:} The JWST data presented in this paper were obtained from the Mikulski Archive for Space Telescopes (MAST) at the Space Telescope Science Institute. The specific observations analyzed can be accessed via \dataset[doi:10.17909/1hrx-cn31]{https://doi.org/doi:10.17909/1hrx-cn31}. The transmission spectra of TOI-270~d reported in this work are available on the Open Science Framework at  
\dataset[https://osf.io/2czbh]{https://osf.io/2czbh}.

\facilities{JWST (NIRISS, NIRSpec, and MIRI)}\\

\bibliography{ms.bib,references.bib}
\bibliographystyle{aasjournal}

\appendix

\section{Data Reduction Robustness Checks} \label{app:robustness}

\begin{figure}
	\includegraphics[width=1\textwidth]{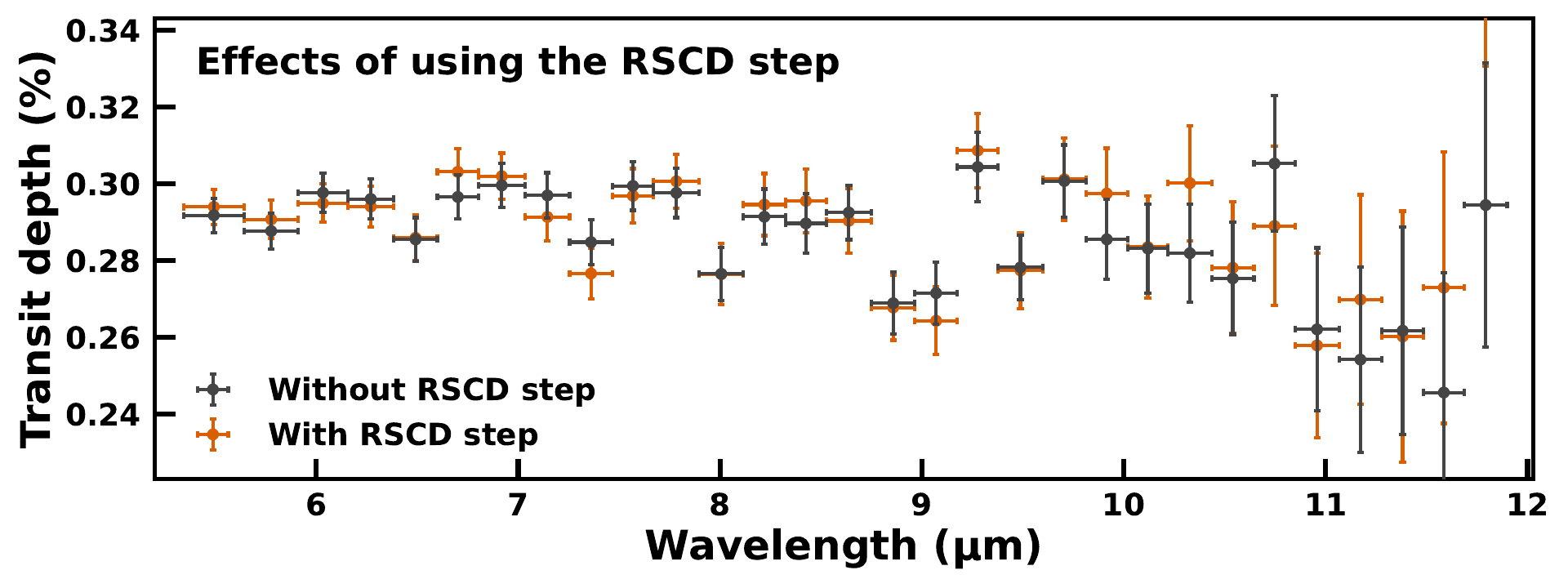}
    \caption{The effect of the RSCD step on the transmission spectrum.
    }
    \label{fig:RSCD_comparison} 
\end{figure}

We conduct a series of robustness checks to test the sensitivity of the MIRI transmission spectrum of TOI-270~d to different data reduction assumptions. First, given that our observation only has 17 groups per integration, we assess the effect of using the standard RSCD (Reset Switch Charge Decay) step. This step flags the first few groups and removes these from the ramp fitting to minimize the influence of the RSCD effect. However, for an observation with only 17 groups, an argument could be made that it is beneficial to use all the data to improve the signal-to-noise ratio. We show the effect on the transmission spectrum in Figure~\ref{fig:RSCD_comparison}. We find that most data points agree within the uncertainties. The benefit of not masking the first few groups is that the uncertainties are smaller when using all the data, particularly at longer wavelengths, where the uncertainty is reduced by 15\%. At the blue end, the uncertainties do not significantly change. On the other hand, it is unclear whether the RSCD effect can cause potential biases in the transmission spectrum when the RSCD step is turned off. Recent attempts to correct for the RSCD and non-linearity effects in a MIRI LRS observation of K2-18~b demonstrated that the transmission spectrum was insensitive to the treatment of these effects \citep{Madhusudhan2025}. To be cautious, we run atmospheric retrievals on both versions of the TOI-270~d spectrum -- with and without the RSCD step, as described in Section~\ref{sec:canonical}. More work is needed to determine the optimal treatment of the RSCD effect.

\begin{figure}
	\includegraphics[width=1\textwidth]{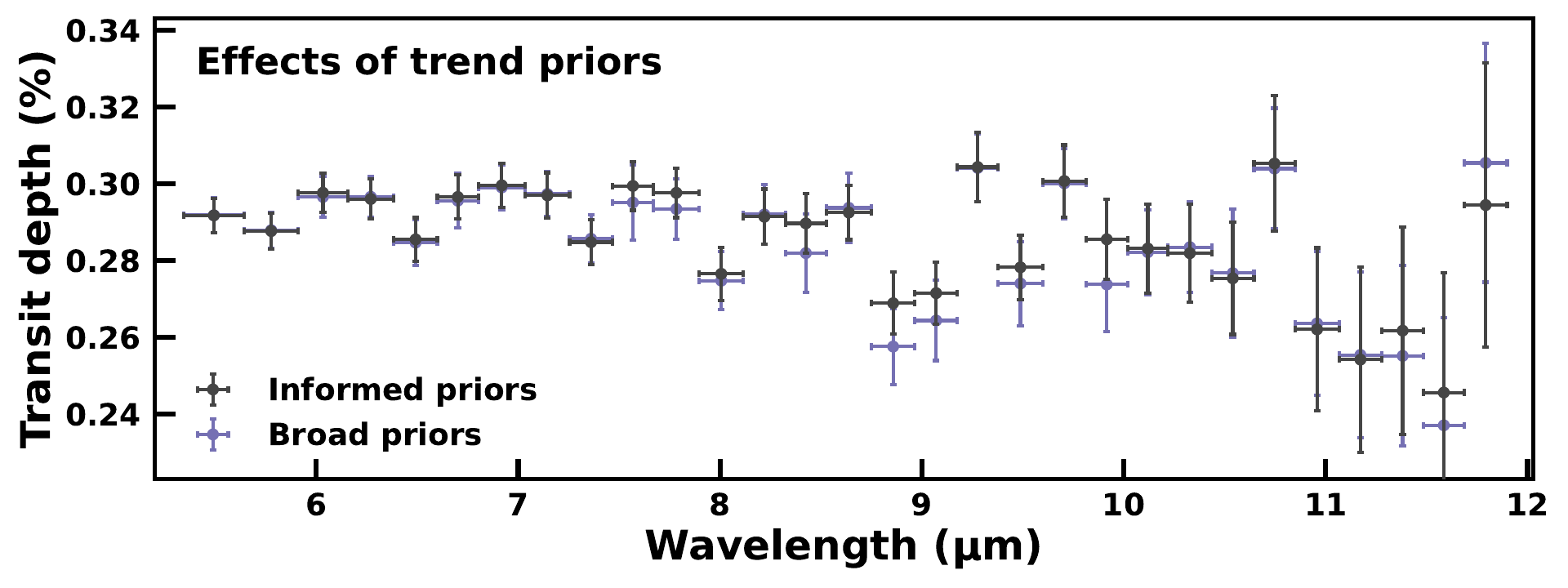}
\caption{Comparison between different prior assumptions for the detector settling timescale. The informed prior case uses the relation outlined in Section~\ref{sec:time-scale}, while the broad prior case uses a broad log uniform prior for the timescale, as described in Appendix~\ref{app:robustness}.
    }
    \label{fig:trend_comparison} 
\end{figure}

Next, we test the effects of applying different priors on the detector settling timescale. This involves using the prior derived from the linear relationship outlined in Section~\ref{sec:time-scale} and comparing this to using a log-uniform prior. For the latter, we use a prior distribution of $\mathcal{U}(-1, -3)$ for the logarithm of the timescale $\tau$ (in days). Figure~\ref{fig:trend_comparison} shows the comparison between these two approaches. We find that the spectra are consistent. The largest difference is that the transit depth uncertainties are smaller when using the informed prior, an average of 11\% smaller below 10~$\mu$m and up to 25\% between 7-9~$\mu$m. For this reason, we adopt the spectrum which uses the informed priors. This method could be used to increase the precision of other MIRI LRS data sets.

\begin{figure}
	\includegraphics[width=1\textwidth]{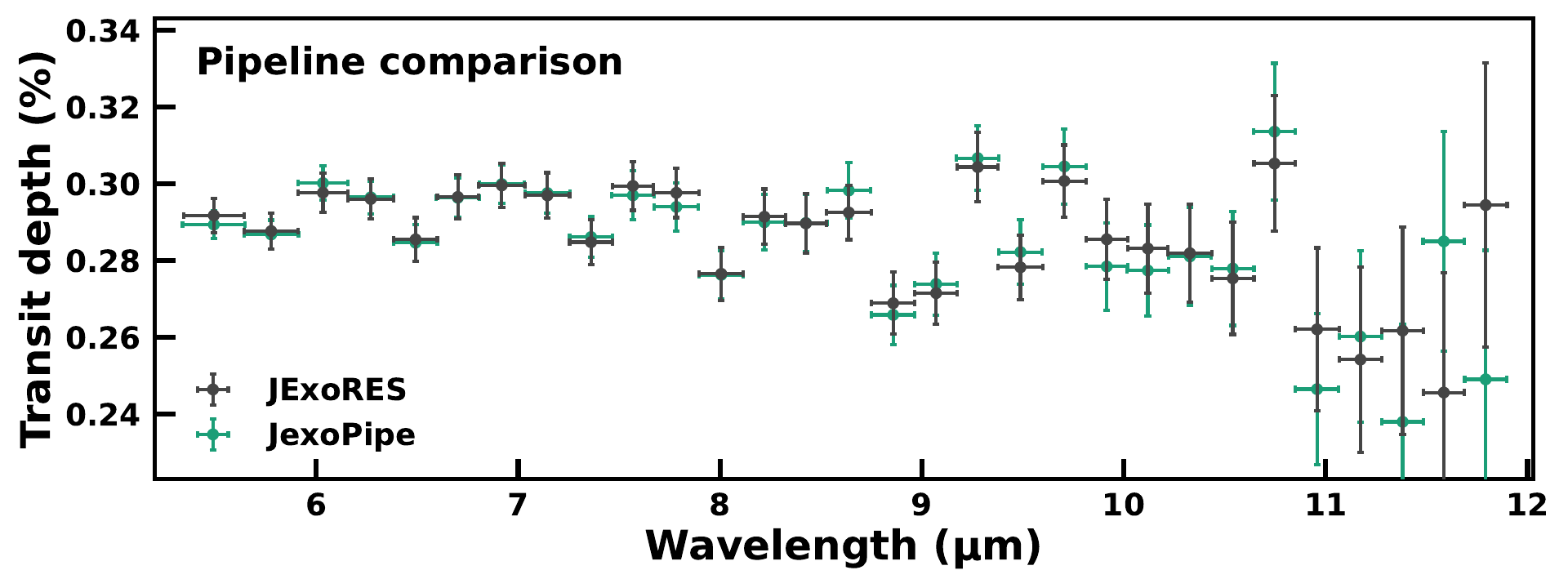}
    \caption{ Comparison between different data reduction
pipelines. The spectra from the \texttt{JExoRES} and \texttt{JexoPipe} pipelines are shown in grey and green, respectively, and are described in Section~\ref{sec:obs}.  Both the default \texttt{JExoRES} spectrum shown and the \texttt{JexoPipe} spectrum omit the RSCD step.
 }
    \label{fig:pipeline_comparison} 
\end{figure}

Finally, we compare the \texttt{JExoRES} and \texttt{JexoPipe} transmission spectra. We show the spectra of both data reduction pipelines in Figure~\ref{fig:pipeline_comparison}. For \texttt{JExoRES}, we use the version without the RSCD step. 
The spectra of both pipelines mostly agree.
For robustness, we use both data reduction pipelines in this work when performing atmospheric retrievals.

\section{Canonical retrieval results}

\begin{figure*}
\centering
	\includegraphics[width=1\textwidth]{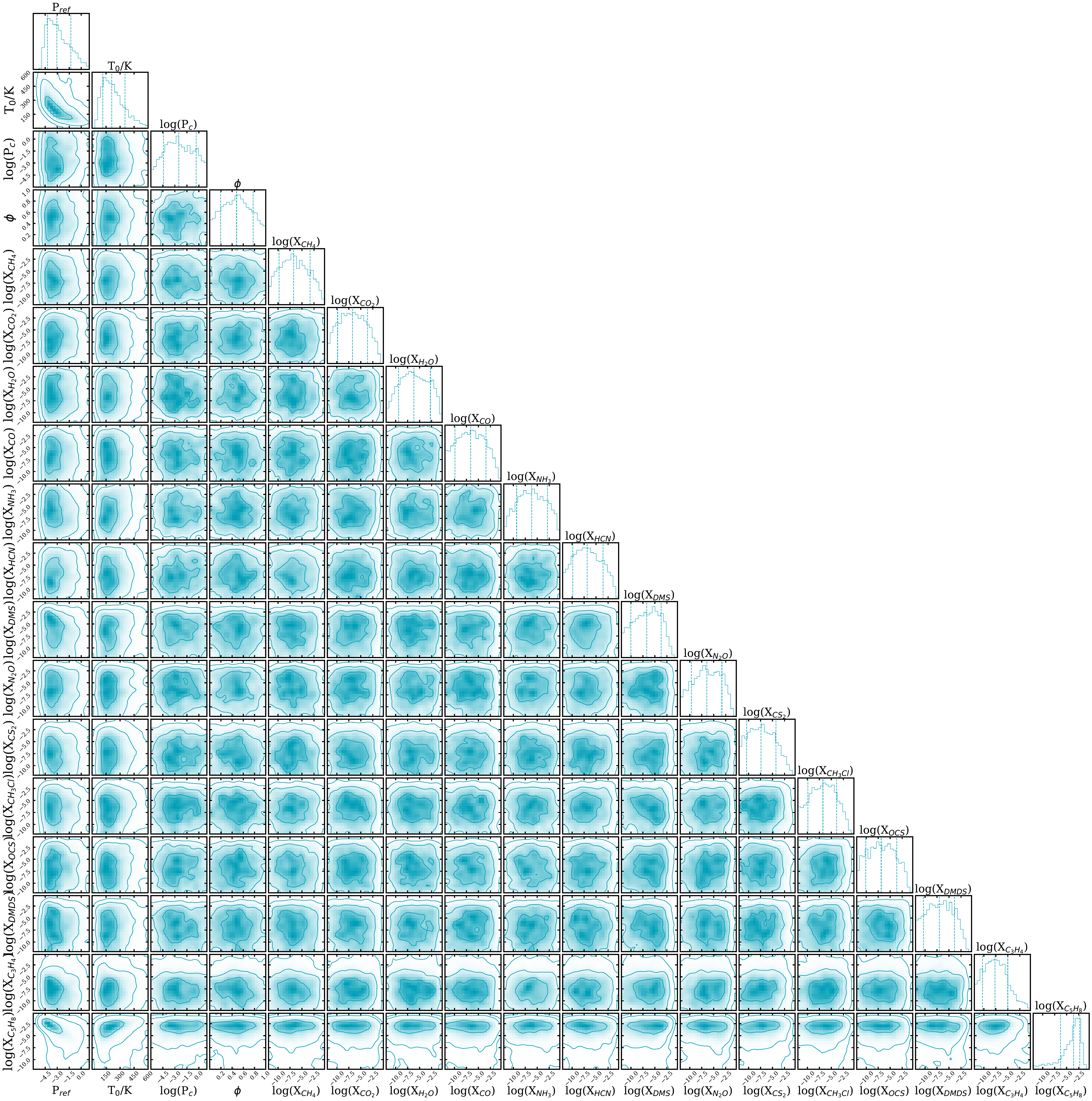}
    \caption{Posterior distributions from the canonical retrieval using the MIRI \texttt{JExoRES} (no RSCD step) data, as described in Section~\ref{sec:canonical}. }
    \label{fig:corner-plot_MIRI} 
\end{figure*}

As outlined in Section~\ref{sec:canonical}, we conduct three sets of canonical retrievals using the current MIRI observations. We present the prior distributions and posterior estimates for all parameters in Table~\ref{tab:retrieval_priors_canonical}. We also present the corner plot for the MIRI \texttt{JExoRES} in Figure~\ref{fig:corner-plot_MIRI}. Note that all abundance constraints have been converted to volume mixing ratios to facilitate comparison with prior studies, while the priors are expressed in terms of mass mixing ratios (which is the default parameterization used in \texttt{pRT}).

\section{Exploration of molecules}

\begin{figure*}
\centering
	\includegraphics[width=0.9\textwidth]{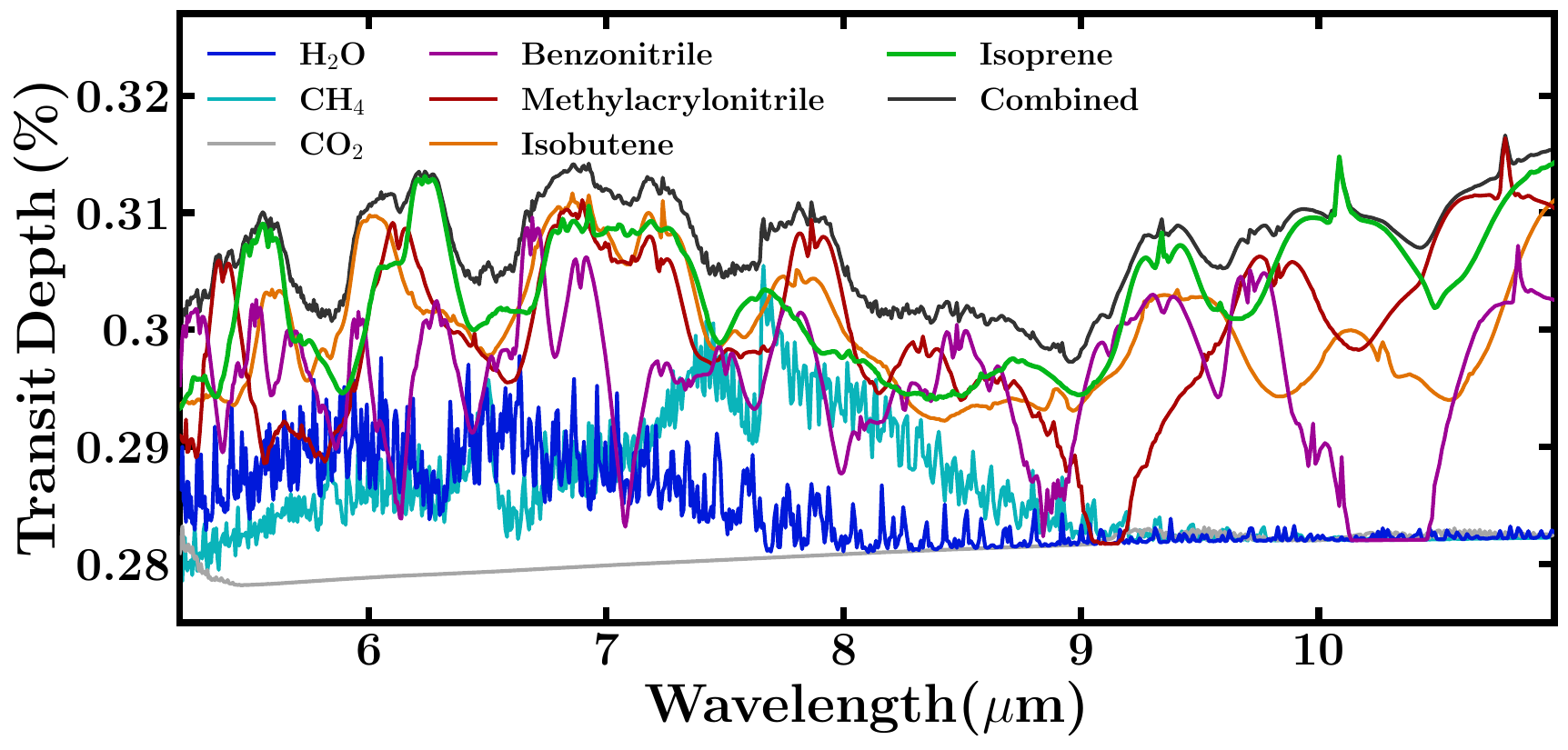}
    \caption{
    Spectral contributions of the 3 base molecular species, the 3 reaching ln~$B$ $\geq$ 2 with MIRI, and isoprene (as in Figure~\ref{fig:fit_plot_MIRI}), in the 5.2-11 $\mu$m range. The different curves show individual contributions from different molecules to a nominal model transmission spectrum of TOI-270~d shown in black and denoted as Combined. The model assumes a volume mixing ratio of 10$^{-2}$ for CH$_4$ and H$_2$O; 10$^{-3}$ for CO$_2$; and 10$^{-4}$ for the other species, consistent with our retrieval estimates. The reference pressure at R$_p$ is 10$^{-2}$~bar, the isothermal temperature profile is set at 260~K, the cloud deck pressure at 10$^{-2}$ bar, and the cloud coverage fraction at 0.4. Each curve corresponds to a transmission spectrum with opacity contributions from a single molecule at a time, in addition to H$_2$-H$_2$ and H$_2$-He collision-induced absorption.
    }
    \label{fig:contribution_plot_MIRI} 
\end{figure*}

\begin{figure*}
\centering
	\includegraphics[width=0.95\textwidth]{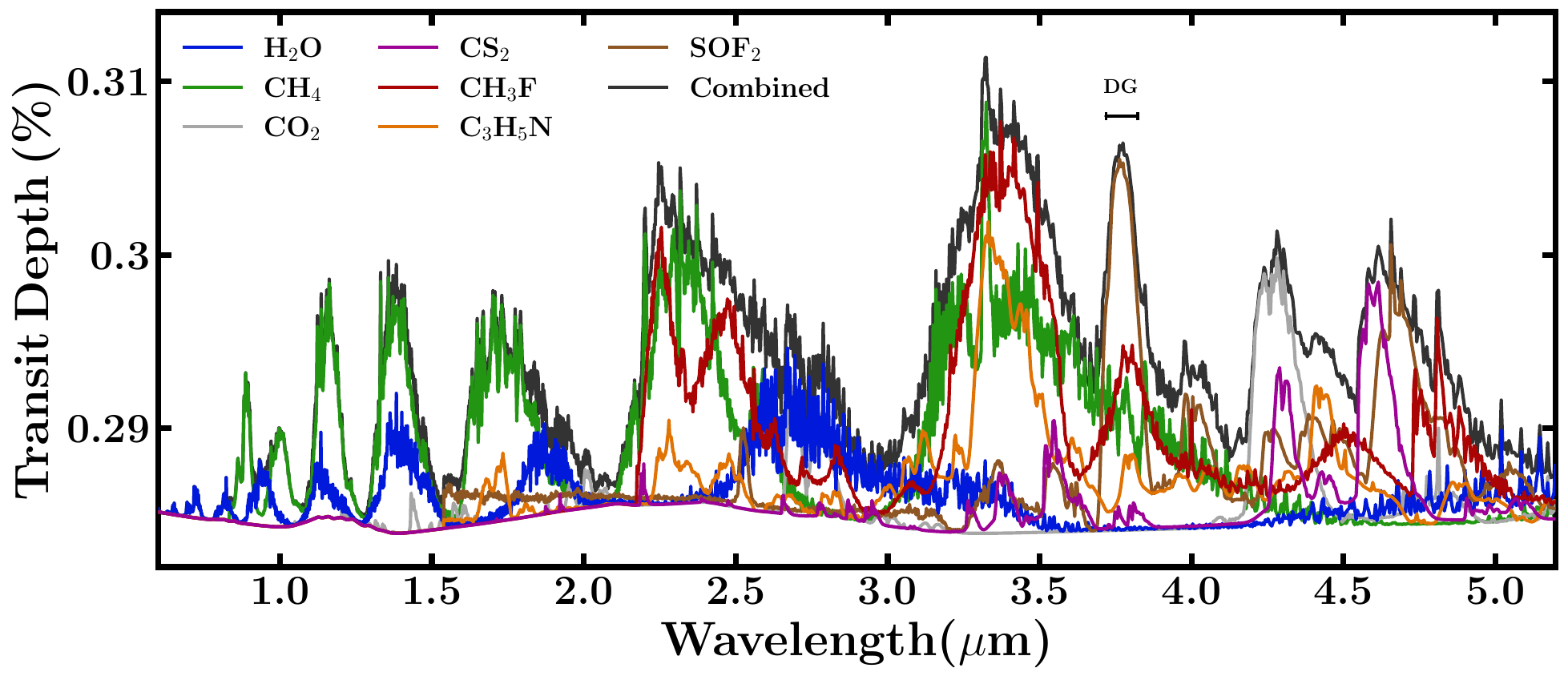}
    \caption{
    Spectral contributions of the 3 base molecular species, and the 4 reaching ln~$B$ $\geq$ 3 with NIRISS+NIRSpec, in the 0.6-5.2 $\mu$m range. The different curves show individual contributions from different molecules to a nominal model transmission spectrum of TOI-270~d shown in black and denoted as Combined. The model assumes a volume mixing ratio of 10$^{-2}$ for CH$_4$ and H$_2$O; 10$^{-3}$ for CO$_2$, CS$_2$, CH$_3$F and SOF$_2$; and 10$^{-5}$ for C$_3$H$_5$N, consistent with our retrieval estimates. 
    Again, P$_{\mathrm{ref}}$~=~10$^{-2}$~bar and T$_{\mathrm{iso}}$~=~260~K, but the cloud deck pressure is 10$^{-2.5}$~bar and the cloud fraction is 0.85. Each curve corresponds to a transmission spectrum with opacity contributions from a single molecule at a time, in addition to H$_2$-H$_2$ and H$_2$-He collision-induced absorption. We also show the the NIRSpec detector gap from 3.72 to 3.82~nm, labeled DG. Since SOF$_2$ has a strong feature in this range and the other species don't, observing the transit depth there would help distinguish between them.
    }
    \label{fig:contribution_plot_NIR} 
\end{figure*}

\begin{figure*}
\centering
	\includegraphics[width=1\textwidth]{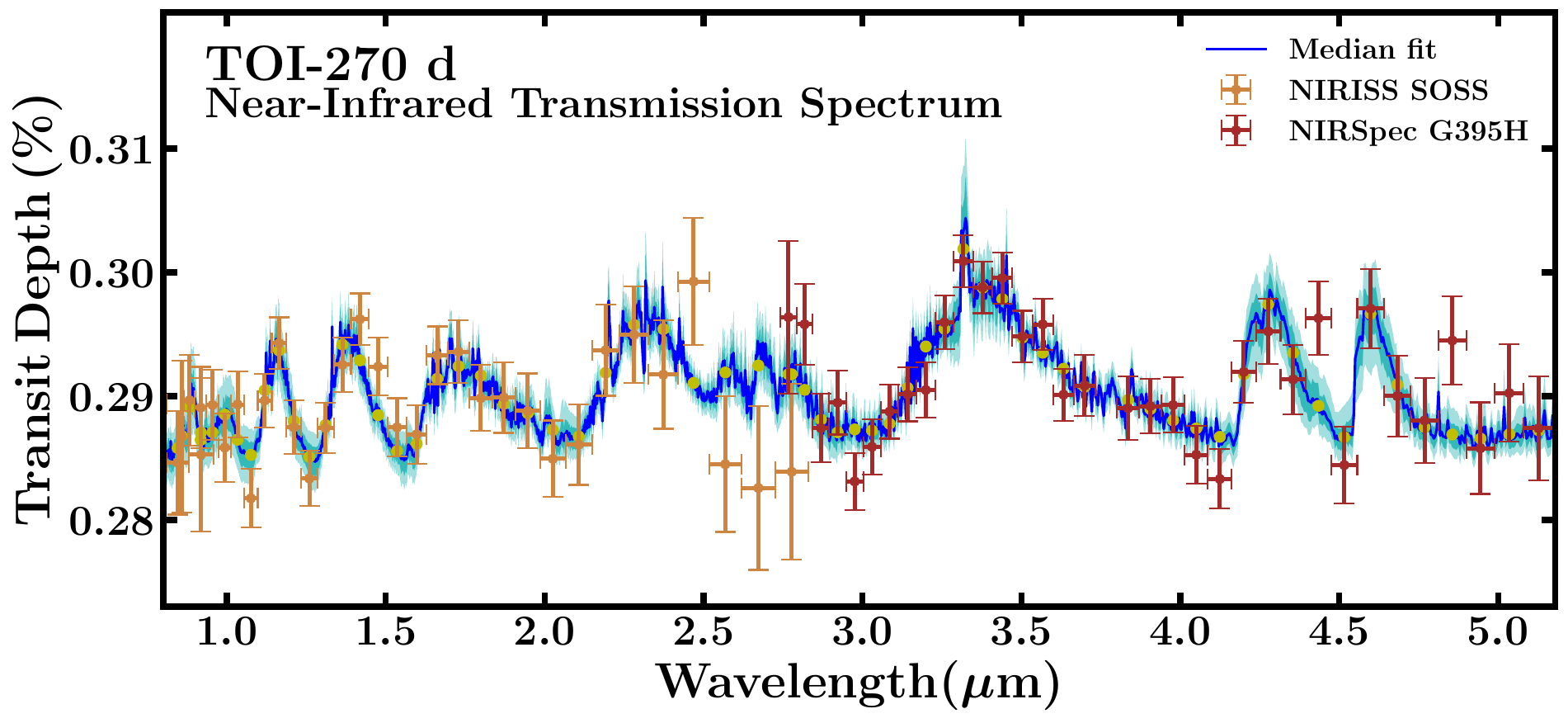}
    \caption{The near-infrared transmission spectrum of TOI-270 d obtained with the JWST NIRISS SOSS and NIRSPEC G395H instruments \citep{Holmberg2024, Constantinou2025}. The model fit corresponds to the 4+C$_2$H$_6$ case, as described in Section~\ref{sec:exploration}. We choose to show this case even though CH$_3$F and C$_3$H$_5$N achieve higher $\ln B$ values in the 4+X exploration (see Table~\ref{tab:all-molecules-NIR}), because C$_2$H$_6$ is a  simpler and more plausible molecule, and was also inferred in \cite{Constantinou2025}.
    The data points with error bars show the observed spectrum. The dark blue curve denotes the median retrieved spectral fit, while the two lighter shaded regions denote the 1$\sigma$ and 2$\sigma$ intervals. The NIRISS data has been offsetted by +32 ppm.
    }
    \label{fig:canonical_plot_NIR} 
\end{figure*}

Here we provide results from the full exploration of molecules, as described in Section~\ref{sec:exploration}. Tables~\ref{tab:all-molecules} and \ref{tab:all-molecules-NIR} contain the evidence for all 203 molecules for the mid-infrared and near-infrared, respectively. We also show the contributions of the most significant molecules in Figures~\ref{fig:contribution_plot_MIRI} and~\ref{fig:contribution_plot_NIR}, along with the NIR spectral fit of the 4+C$_2$H$_6$ model in Figure~\ref{fig:canonical_plot_NIR} for completeness.

\input{first-sweep_MIRI-cloudy-new}
\input{first-sweep_NIR-cloudy}

\section{Remaining molecules}
\label{sec:remaining-molecules}

Here we provide the molecular species not listed in Tables~\ref{tab:all-molecules} and \ref{tab:all-molecules-NIR}: 1-Butyne (C\textsubscript{4}H\textsubscript{6}), Acrylonitrile (C\textsubscript{3}H\textsubscript{3}N), Allene (C\textsubscript{3}H\textsubscript{4}), Ammonia (NH\textsubscript{3}), Benzene (C\textsubscript{6}H\textsubscript{6}), Diisopropylamine (C\textsubscript{6}H\textsubscript{15}N), Ethylene (C\textsubscript{2}H\textsubscript{4}), Ethylenediamine (C\textsubscript{2}H\textsubscript{8}N\textsubscript{2}), Formaldehyde (CH\textsubscript{2}O), Hydrazine (N\textsubscript{2}H\textsubscript{4}), Imidogen (NH), Methylhydrazine (CH\textsubscript{6}N\textsubscript{2}), n-Tridecane (C\textsubscript{13}H\textsubscript{28}), Naphthalene (C\textsubscript{10}H\textsubscript{8}), Nitrous acid (HNO\textsubscript{2}), Sulfur Trioxyde (SO\textsubscript{3}), Sulfur-Dioxide (SO\textsubscript{2}), Trifluoronitrosomethane (CF\textsubscript{3}NO), Trimethylamine (C\textsubscript{3}H\textsubscript{9}N)

\end{document}

%% file: first-sweep_MIRI-cloudy-new.tex
\clearpage 
\startlongtable
\begin{deluxetable*}{lllll}
  \tablecaption{Model preferences for 203 molecules considered in this work to explain the MIRI data. These were explored in the retrievals using \texttt{pRT} for the two JWST MIRI reductions: \texttt{JExoRES} (without RCSD step) and \texttt{JexoPipe}. Log Bayes factors ($\ln B$) are given for individual molecules, along with $\sigma$ values in brackets according to the conversion from \cite{Sellke_Bayarri_2001}, whenever ln~$B$ $\geq$ 1; but cf. \citet{Thorngren_Sing_2025}. The $\ln B$ is for the model with the relevant molecule included, versus the baseline model only including CH$_4$, CO$_2$, and H$_2$O, as discussed in Section \ref{sec:exploration}. In all the retrievals we allow for the presence of patchy clouds in the models, even though there is no significant evidence. This is done to obtain conservative values of $\ln B$, as shown in Table~\ref{tab:exploration-MIRI} for the top few molecules. We also provide the log-evidence and reduced chi-squared $\chi_r^2$ for the baseline model in the first row, where $\chi_r^2 := \frac{\chi^2}{N_p-n_c}$, where $N_p = 26$ is the number of data points and $n_c = 3$ is the approximate number of constrained parameters: $P_{\mathrm{ref}}$, $T_{\mathrm{iso}}$ and log(X$_{H_2O}$) in this case. We find that the \texttt{JExoRES} spectrum is better fitted by the baseline compared to the \texttt{JexoPipe} spectrum, which is why it needs less the presence of a fourth molecule, thus showing somewhat lower model preferences for their addition. For all the species with $\ln B \geq 1$, we also conduct retrievals with narrower priors on CH$_4$, CO$_2$ and H$_2$O abundances based on NIR constraints (see \ref{sec:exploration}). The species with $\ln B \leq 0$ in the \texttt{JExoRES} spectrum are not shown here.
  \label{tab:all-molecules}}
  \tablehead{
    \colhead{Molecule} &
    \colhead{MIRI \texttt{JExoRES}} &
    \colhead{+NIR Priors} &
    \colhead{MIRI \texttt{JexoPipe}} &
    \colhead{+NIR Priors}
  }
  \startdata 
Baseline ln~$Z$ ($\chi_r^2$) & 199.31 (1.49) & 199.53 (1.54) & 193.12 (2.17) & 192.87 (2.25) \\ \hline
Benzonitrile (C$_7$H$_5$N)                     & 3.43 (3.1$\sigma$)         & 2.79 (2.9$\sigma$) & 5.25 (3.7$\sigma$)           & 4.67 (3.5$\sigma$)\\                                            
Methylacrylonitrile (C$_4$H$_5$N)              & 2.48 (2.7$\sigma$)         & 2.25 (2.6$\sigma$) & 3.22 (3.0$\sigma$)           & 3.07 (3.0$\sigma$)\\                                  
Isobutene (C$_4$H$_8$)                         & 2.32 (2.7$\sigma$)         & 1.79 (2.4$\sigma$) & 4.27 (3.4$\sigma$)           & 3.95 (3.3$\sigma$)\\                                 
Styrene (C$_8$H$_8$)                           & 1.71 (2.4$\sigma$)         & 1.39 (2.3$\sigma$) & 1.76 (2.4$\sigma$)           & 1.77 (2.4$\sigma$)\\           
Cumene (C$_9$H$_{12}$)                         & 1.40 (2.3$\sigma$)         & 1.05 (2.1$\sigma$) & 2.90 (2.9$\sigma$)           & 2.34 (2.7$\sigma$)\\         
Isoprene (C$_5$H$_8$)                          & 1.34 (2.2$\sigma$)         & 1.28 (2.2$\sigma$) & 2.04 (2.6$\sigma$)           & 1.90 (2.5$\sigma$)\\        
2-Methyl-1-butene (C$_5$H$_{10}$)              & 1.06 (2.1$\sigma$)         & 0.78 (\textless 2$\sigma$) & 2.56 (2.8$\sigma$)   & 2.33 (2.3$\sigma$)\\     
Pyridine (C$_5$H$_5$N)                         & 1.02 (2.0$\sigma$)         & 0.93 (\textless 2$\sigma$) & 2.19 (2.6$\sigma$)   & 2.47 (2.7$\sigma$)\\     
Propene (C$_3$H$_6$)                           & 1.00 (2.0$\sigma$)         & 1.30 (2.2$\sigma$) & 2.61 (2.8$\sigma$)           & 2.53 (2.8$\sigma$)\\     
Ethyl cyanide (C$_3$H$_5$N)                    & 0.96 (\textless 2$\sigma$) & & 1.90 (2.5$\sigma$) & \\                                                   
tert-Butylbenzene (C$_{10}$H$_{14}$)           & 0.95 (\textless 2$\sigma$) & & 1.42 (2.3$\sigma$) & \\                                                       
2,4,4-Trimethyl-1-pentene (C$_8$H$_{16}$)        & 0.93 (\textless 2$\sigma$) & & 1.87 (2.5$\sigma$) & \\                                                     
Ethyl benzene (C$_8$H$_{10}$)                  & 0.88 (\textless 2$\sigma$) & & 1.67 (2.4$\sigma$) & \\                                                     
4-Picoline (C$_6$H$_7$N)                       & 0.86 (\textless 2$\sigma$) & & 2.36 (2.7$\sigma$) & \\                                                   
2-Methyl-1-pentene (C$_6$H$_{12}$)             & 0.85 (\textless 2$\sigma$) & & 2.19 (2.6$\sigma$) & \\                                                     
Aniline (C$_6$H$_7$N)                          & 0.77 (\textless 2$\sigma$) & & 0.71 (\textless 2$\sigma$)& \\                                                   
Dimethyl disulfide (C$_2$H$_6$S$_2$)           & 0.62 (\textless 2$\sigma$) & & 1.02 (2.0$\sigma$) & \\                                                       
2,4,4-Trimethyl-2-pentene (C$_8$H$_{16}$)        & 0.61 (\textless 2$\sigma$) & & 1.90 (2.5$\sigma$) & \\                                                                              
Amyl nitrate (C$_5$H$_{11}$NO$_3$)             & 0.56 (\textless 2$\sigma$) & & 1.04 (2.0$\sigma$) & \\                                                                 
Isocumene (C$_9$H$_{12}$)                      & 0.52 (\textless 2$\sigma$) & & 1.00 (2.0$\sigma$) & \\                                                                              
D-Limonene (C$_{10}$H$_{16}$)                  & 0.51 (\textless 2$\sigma$) & & 2.34 (2.7$\sigma$) & \\                                                                                
Toluene (C$_7$H$_8$)                           & 0.47 (\textless 2$\sigma$) & & 0.89 (\textless 2$\sigma$)& \\                                                                            
1,1-Dimethylhydrazine (C$_2$H$_8$N$_2$)         & 0.46 (\textless 2$\sigma$) & & 1.05 (2.0$\sigma$) & \\                                                             
cis-2-Pentene (C$_5$H$_{10}$)                  & 0.45 (\textless 2$\sigma$) & & 1.34 (2.2$\sigma$) & \\                                                                              
Pentane (C$_5$H$_{12}$)                        & 0.44 (\textless 2$\sigma$) & & 0.98 (\textless 2$\sigma$)& \\                                                                                 
Dimethyl sulfide (C$_2$H$_6$S)                 & 0.43 (\textless 2$\sigma$) & & 0.39 (\textless 2$\sigma$)& \\                          
3-Methyl-1-butene (C$_5$H$_{10}$)              & 0.41 (\textless 2$\sigma$) & & 1.37 (2.2$\sigma$) & \\                                                                               
1-Pentene (C$_5$H$_{10}$)                      & 0.40 (\textless 2$\sigma$) & & 1.15 (2.1$\sigma$) & \\                                                                              
1-Butene (C$_4$H$_8$)                          & 0.40 (\textless 2$\sigma$) & & 0.96 (\textless 2$\sigma$)& \\                                                                            
3-Ethyltoluene (C$_9$H$_{12}$)                 & 0.40 (\textless 2$\sigma$) & & 0.67 (\textless 2$\sigma$)& \\                                                                              
3-Carene (C$_{10}$H$_{14}$)                    & 0.37 (\textless 2$\sigma$) & & 0.85 (\textless 2$\sigma$)& \\                                                                                
sec-Butylbenzene (C$_{10}$H$_{14}$)            & 0.36 (\textless 2$\sigma$) & & 0.92 (\textless 2$\sigma$)& \\                                                                                
Isobutane (C$_4$H$_{10}$)                      & 0.36 (\textless 2$\sigma$) & & 0.63 (\textless 2$\sigma$)& \\                                                                              
Heptane (C$_7$H$_{16}$)                        & 0.35 (\textless 2$\sigma$) & & 0.53 (\textless 2$\sigma$)&  \\                                                                              
o-Toluidine (C$_7$H$_9$N)                      & 0.35 (\textless 2$\sigma$) & & 1.20 (2.1$\sigma$) & \\                     
2-Butene (C$_4$H$_8$)                          & 0.32 (\textless 2$\sigma$) & & 1.06 (2.1$\sigma$) & \\                                                                               
Methylamine (CH$_5$N)                          & 0.32 (\textless 2$\sigma$) & & 2.47 (2.7$\sigma$) & \\                 
2,6-Diethylaniline (C$_{10}$H$_{15}$N)          & 0.31 (\textless 2$\sigma$) & & 1.45 (2.3$\sigma$) & \\                                 
Ethane (C$_2$H$_6$)                            & 0.31 (\textless 2$\sigma$) & & 0.22 (\textless 2$\sigma$)& \\                                                                             
Isohexane (C$_6$H$_{14}$)                      & 0.29 (\textless 2$\sigma$) & & 0.56 (\textless 2$\sigma$)& \\                                                                               
3-Methylhexane (C$_7$H$_{16}$)                 & 0.27 (\textless 2$\sigma$) & & 0.53 (\textless 2$\sigma$)& \\                                                                               
Cycloheptene (C$_7$H$_{12}$)                   & 0.26 (\textless 2$\sigma$) & & 0.84 (\textless 2$\sigma$)& \\                                                                               
2,4-Dimethylpentane (C$_7$H$_{16}$)             & 0.25 (\textless 2$\sigma$) & & 0.48 (\textless 2$\sigma$)& \\                                                                               
Hydrogen cyanide (HCN)                         & 0.25 (\textless 2$\sigma$) & & 0.26 (\textless 2$\sigma$)& \\                 
Chloroethane (C$_2$H$_5$Cl)                    & 0.24 (\textless 2$\sigma$) & & 1.04 (2.0$\sigma$) & \\                         
1,2,3-Trimethylbenzene (C$_9$H$_{12}$)           & 0.24 (\textless 2$\sigma$) & & 0.63 (\textless 2$\sigma$)& \\                                                                               
1,2,3,5-Tetramethylbenzene (C$_{10}$H$_{14}$)     & 0.24 (\textless 2$\sigma$) & & 0.02 (\textless 2$\sigma$) & \\                                                                              
1,2,3,4-Tetramethylbenzene (C$_{10}$H$_{14}$)     & 0.23 (\textless 2$\sigma$) & & 0.42 (\textless 2$\sigma$)& \\                                                                              
2-Ethyltoluene (C$_9$H$_{12}$)                 & 0.23 (\textless 2$\sigma$) & & 0.49 (\textless 2$\sigma$)& \\                                                                               
Ethylamine (C$_2$H$_7$N)                       & 0.23 (\textless 2$\sigma$) & & 2.26 (2.7$\sigma$) & \\                      
13-Butadiene (C$_4$H$_6$)                      & 0.22 (\textless 2$\sigma$) & & -0.13 (-)           & \\                                                                             
1-Heptene (C$_7$H$_{14}$)                      & 0.21 (\textless 2$\sigma$) & & 0.42 (\textless 2$\sigma$)& \\                                                                               
Propylene-sulfide (C$_3$H$_6$S)                & 0.19 (\textless 2$\sigma$) & & 0.70 (\textless 2$\sigma$)& \\                             
Butane (C$_4$H$_{10}$)                         & 0.18 (\textless 2$\sigma$) & & 0.44 (\textless 2$\sigma$)& \\                                                                                
Acetonitrile (C$_2$H$_3$N)                     & 0.18 (\textless 2$\sigma$) & & -0.09 (-)           & \\                     
Vinyl toluene (C$_9$H$_{10}$)                  & 0.17 (\textless 2$\sigma$) & & 0.41 (\textless 2$\sigma$)& \\                                                                                 
Cyclopentene (C$_5$H$_8$)                      & 0.16 (\textless 2$\sigma$) & & 0.71 (\textless 2$\sigma$)& \\                                                                              
Propane (C$_3$H$_8$)                           & 0.16 (\textless 2$\sigma$) & & 0.44 (\textless 2$\sigma$)& \\                                                                             
Allylamine (C$_3$H$_7$N)                       & 0.16 (\textless 2$\sigma$) & & 0.41 (\textless 2$\sigma$)& \\                     
Benzenethiol (C$_6$H$_6$S)                     & 0.15 (\textless 2$\sigma$) & & 0.03 (\textless 2$\sigma$)    & \\                     
Isopentane (C$_5$H$_{12}$)                     & 0.14 (\textless 2$\sigma$) & & 0.11 (\textless 2$\sigma$)& \\                                                                                
Diethyl sulfide (C$_4$H$_{10}$S)               & 0.13 (\textless 2$\sigma$) & & 0.45 (\textless 2$\sigma$)& \\                             
2,2-Dimethylbutane (C$_6$H$_{14}$)              & 0.13 (\textless 2$\sigma$) & & 0.19 (\textless 2$\sigma$)& \\                                                                                
N,N-Diethylaniline (C$_{10}$H$_{15}$N)          & 0.13 (\textless 2$\sigma$) & & -0.19 (-)           & \\                                 
Myrcene (C$_{10}$H$_{16}$)                     & 0.12 (\textless 2$\sigma$) & & 0.72 (\textless 2$\sigma$)& \\                                                                                  
Cyclopentane (C$_5$H$_{10}$)                   & 0.10 (\textless 2$\sigma$) & & 0.61 (\textless 2$\sigma$)& \\
2-Vinylpyridine (C$_7$H$_7$N)                  & 0.10 (\textless 2$\sigma$) & & -0.28 (-)           & \\                                                                                 
Phosphorus mononitride (PN)                    & 0.10 (\textless 2$\sigma$) & & -0.40 (-)           & \\                                                                             
Cadaverine (C$_5$H$_{14}$N$_2$)                & 0.09 (\textless 2$\sigma$) & & 0.70 (\textless 2$\sigma$)& \\    
Nitromethane (CH$_3$NO$_2$)                    & 0.09 (\textless 2$\sigma$) & & 0.68 (\textless 2$\sigma$)& \\                                                          
Ethyne (C$_2$H$_2$)                            & 0.09 (\textless 2$\sigma$) & & -0.22 (-)           & \\                                                      
Hexane (C$_6$H$_{14}$)                         & 0.08 (\textless 2$\sigma$) & & 0.48 (\textless 2$\sigma$)& \\                                                            
Octane (C$_8$H$_{18}$)                         & 0.08 (\textless 2$\sigma$) & & 0.11 (\textless 2$\sigma$)& \\      
Fluoromethane (CH$_3$F)                        & 0.08 (\textless 2$\sigma$) & & 0.02 (\textless 2$\sigma$) & \\                                                                        
Propyne (C$_3$H$_4$)                           & 0.07 (\textless 2$\sigma$) & & -0.31 (-)           & \\                                             
Carbon monoxyde (CO)                           & 0.07 (\textless 2$\sigma$) & & -0.47 (-)           & \\                              
1-Undecene (C$_{11}$H$_{22}$)                  & 0.06 (\textless 2$\sigma$) & & -0.02 (-)           & \\           
1-Octene (C$_8$H$_{16}$)                       & 0.05 (\textless 2$\sigma$) & & 0.20 (\textless 2$\sigma$)& \\       
n-Tetradecane (C$_{14}$H$_{30}$)               & 0.05 (\textless 2$\sigma$) & & 0.07 (\textless 2$\sigma$)& \\        
tert-Amylamine (C$_5$H$_{13}$N)                & 0.04 (\textless 2$\sigma$) & & 0.24 (\textless 2$\sigma$)& \\                                         
n-Nonane (C$_9$H$_{20}$)                       & 0.04 (\textless 2$\sigma$) & & -0.19 (-)           & \\                   
Quinoline (C$_9$H$_7$N)                        & 0.03 (\textless 2$\sigma$) & & 0.09 (\textless 2$\sigma$)& \\                                 
1-Hexene (C$_6$H$_{12}$)                       & 0.02 (\textless 2$\sigma$) & & 0.22 (\textless 2$\sigma$)& \\      
Carbon Monosulfide (CS)                        & 0.02 (\textless 2$\sigma$) & & -0.02 (-)           & \\                                 
Carbonyl sulfide (OCS)                         & 0.02 (\textless 2$\sigma$) & & -0.30 (-)           & \\                              
trans-2-Pentene (C$_5$H$_{10}$)                & 0.01 (\textless 2$\sigma$) & & 0.28 (\textless 2$\sigma$)& \\   
Cyclodecane (C$_{10}$H$_{20}$)                 & 0.01 (\textless 2$\sigma$) & & 0.18 (\textless 2$\sigma$)& \\              
Hexadecane (C$_{16}$H$_{34}$)                  & 0.01 (\textless 2$\sigma$) & & 0.01 (\textless 2$\sigma$) & \\     
o-Xylene (C$_8$H$_{10}$)                       & 0.00 (\textless 2$\sigma$) & & 0.35 (\textless 2$\sigma$)& \\  
Mesitylene (C$_9$H$_{12}$)                     & 0.00 (\textless 2$\sigma$) & & 0.28 (\textless 2$\sigma$)& \\                                                                                                                                                         
\enddata
\end{deluxetable*}

%% file: first-sweep_NIR-cloudy.tex
\startlongtable
\begin{deluxetable*}{llll}
  \tablecaption{Model preferences for the 203 molecules considered in this work to explain the NIRISS + NIRSpec data. For all the species with $\ln B \geq 2$ and a few others, we also show a run with CH$_4$+CO$_2$+H$_2$O+CS$_2$+X ("4+X" hereafter). The species with $\ln B \leq 0$ spectrum are not shown.
  If a species is shown neither here nor in Table~\ref{tab:all-molecules}, we list it in Appendix~\ref{sec:remaining-molecules}.
  }
  \label{tab:all-molecules-NIR}
  \tablehead{
    \colhead{Molecule} &
    \colhead{NIR 3+X} &
    \colhead{NIR 4+X}
  }
  \startdata
Baseline ln~$Z$ ($\chi_r^2$) & 23695.28 (0.98) & 23701.15 (0.97)\\ \hline
Carbon disulfide (CS\textsubscript{2}) & 5.87 (3.9$\sigma$) & N/A \\
Fluoromethane (CH\textsubscript{3}F) & 4.44 (3.4$\sigma$) & 2.99 (2.9$\sigma$) \\
Ethyl cyanide (C\textsubscript{3}H\textsubscript{5}N) & 3.39 (3.1$\sigma$) & 2.53 (2.8$\sigma$) \\
Thionyl fluoride (SOF\textsubscript{2}) & 3.04 (3.0$\sigma$) & 0.03 (\textless 2$\sigma$)\\
Hexamethylphosphoramide (C\textsubscript{6}H\textsubscript{18}N\textsubscript{3}OP) & 2.89 (2.9$\sigma$) & 0.70 (\textless 2$\sigma$) \\
Dimethyl sulfoxide (C\textsubscript{2}H\textsubscript{6}OS) & 2.70 (2.8$\sigma$) & 0.38 (\textless 2$\sigma$) \\
N,N Dimethylformamide (C\textsubscript{3}H\textsubscript{7}NO) & 2.61 (2.8$\sigma$) & 0.90 (\textless 2$\sigma$) \\
Dimethyl sulfate (C\textsubscript{2}H\textsubscript{6}O\textsubscript{4}S) & 2.58 (2.8$\sigma$) & 0.32 (\textless 2$\sigma$) \\
Dimethyl disulfide (C\textsubscript{2}H\textsubscript{6}S\textsubscript{2}) & 2.29 (2.7$\sigma$) & 1.25 (2.2$\sigma$) \\
Allyl isothiocyanate (C\textsubscript{4}H\textsubscript{5}NS) & 2.09 (2.6$\sigma$) & 0.06 (\textless 2$\sigma$) \\
Diethyl sulfate (C\textsubscript{4}H\textsubscript{10}O\textsubscript{4}S) & 2.05 (2.6$\sigma$) & 1.31 (2.2$\sigma$) \\
Cycloheptene (C\textsubscript{7}H\textsubscript{12}) & 1.81 (2.5$\sigma$) \\
Cyclopentene (C\textsubscript{5}H\textsubscript{8}) & 1.76 (2.4$\sigma$) \\
Cyclohexene (C\textsubscript{6}H\textsubscript{10}) & 1.64 (2.4$\sigma$) \\
Ethane (C\textsubscript{2}H\textsubscript{6}) & 1.61 (2.4$\sigma$) & 1.83 (2.5$\sigma$) \\
Carbon monoxide (CO) & 1.59 (2.3$\sigma$) \\ 
Isocyanic acid (CHNO) & 1.47 (2.3$\sigma$) \\
4-Vinyl-1-cyclohexene (C\textsubscript{8}H\textsubscript{12}) & 1.46 (2.3$\sigma$) \\
Thioformaldehyde (H\textsubscript{2}CS) & 1.44 (2.3$\sigma$) \\ 
Methylamine (CH\textsubscript{5}N) & 1.44 (2.3$\sigma$) \\
Thiophosegene (CCl\textsubscript{2}S) & 1.43 (2.3$\sigma$) \\
(1S)-(-)-$\alpha$-Pinene (C\textsubscript{10}H\textsubscript{16}) & 1.37 (2.2$\sigma$) \\
Propylene sulfide (C\textsubscript{3}H\textsubscript{6}S) & 1.37 (2.2$\sigma$) \\
Dimethylamine (C\textsubscript{2}H\textsubscript{7}N) & 1.34 (2.2$\sigma$) \\
Dichloromethane (CH\textsubscript{2}Cl\textsubscript{2}) & 1.33 (2.2$\sigma$) \\
Cyclodecane (C\textsubscript{10}H\textsubscript{20}) & 1.30 (2.2$\sigma$) \\
Cyclopentane (C\textsubscript{5}H\textsubscript{10}) & 1.17 (2.1$\sigma$) \\
Isobutyronitrile (C\textsubscript{4}H\textsubscript{7}N) & 1.17 (2.1$\sigma$) \\
Propyleneimine (C\textsubscript{3}H\textsubscript{7}N) & 1.11 (2.1$\sigma$) \\
Carbonyl sulfide (OCS) & 1.10 (2.1$\sigma$) \\
Cyclohexane (C\textsubscript{6}H\textsubscript{12}) & 1.08 (2.1$\sigma$) \\
2-Carene (C\textsubscript{10}H\textsubscript{14}) & 1.03 (2.0$\sigma$) \\
Tetralin (C\textsubscript{10}H\textsubscript{12}) & 1.01 (2.0$\sigma$) \\
2-Methyl-1-butene (C\textsubscript{5}H\textsubscript{10}) & 1.00 (2.0$\sigma$) \\
2-Pentene-cis (C\textsubscript{5}H\textsubscript{10}) & 0.90 (\textless 2$\sigma$) \\
3-Carene (C\textsubscript{10}H\textsubscript{14}) & 0.88 (\textless 2$\sigma$) \\
Cycloheptane (C\textsubscript{7}H\textsubscript{14}) & 0.88 (\textless 2$\sigma$) \\
n-Butyl isocyanate (C\textsubscript{5}H\textsubscript{9}NO) & 0.86 (\textless 2$\sigma$) \\
3-Picoline (C\textsubscript{6}H\textsubscript{7}N) & 0.85 (\textless 2$\sigma$) \\
Methanethiol (CH\textsubscript{4}S) & 0.85 (\textless 2$\sigma$) \\
(1S)-(-)-$\beta$-Pinene (C\textsubscript{10}H\textsubscript{16}) & 0.82 (\textless 2$\sigma$) \\
2,4-Diisocyanatotoluene (C\textsubscript{9}H\textsubscript{6}N\textsubscript{2}O\textsubscript{2}) & 0.81 (\textless 2$\sigma$) \\
Silane (SiH\textsubscript{4}) & 0.79 (\textless 2$\sigma$) \\
1-Hexene (C\textsubscript{6}H\textsubscript{12}) & 0.78 (\textless 2$\sigma$) \\
Dimethyl sulfide (C\textsubscript{2}H\textsubscript{6}S) & 0.77 (\textless 2$\sigma$) \\
1-Pentene (C\textsubscript{5}H\textsubscript{10}) & 0.76 (\textless 2$\sigma$) \\
2-Methyl-2-butene (C\textsubscript{5}H\textsubscript{10}) & 0.75 (\textless 2$\sigma$) \\
3-Ethyltoluene (C\textsubscript{9}H\textsubscript{12}) & 0.74 (\textless 2$\sigma$) \\
Cyclohexanethiol (C\textsubscript{6}H\textsubscript{12}S) & 0.73 (\textless 2$\sigma$) \\
Propene (C\textsubscript{3}H\textsubscript{6}) & 0.73 (\textless 2$\sigma$) \\
Dioxygen (O\textsubscript{2}) & 0.72 (\textless 2$\sigma$) \\
trans-2-Pentene (C\textsubscript{5}H\textsubscript{10}) & 0.71 (\textless 2$\sigma$) \\
Heptane (C\textsubscript{7}H\textsubscript{16}) & 0.70 (\textless 2$\sigma$) \\
1-Heptene (C\textsubscript{7}H\textsubscript{14}) & 0.69 (\textless 2$\sigma$) \\
Toluene (C\textsubscript{7}H\textsubscript{8}) & 0.69 (\textless 2$\sigma$) \\
1,2,3,4-Tetramethylbenzene (C\textsubscript{10}H\textsubscript{14}) & 0.67 (\textless 2$\sigma$) \\
Amyl nitrate (C\textsubscript{5}H\textsubscript{11}NO\textsubscript{3}) & 0.67 (\textless 2$\sigma$) \\
Diethylamine (C\textsubscript{4}H\textsubscript{11}N) & 0.67 (\textless 2$\sigma$) \\
Ethylamine (C\textsubscript{2}H\textsubscript{7}N) & 0.65 (\textless 2$\sigma$) \\
m-Xylene (C\textsubscript{8}H\textsubscript{10}) & 0.65 (\textless 2$\sigma$) \\
Thioglycol (C\textsubscript{2}H\textsubscript{6}OS) & 0.64 (\textless 2$\sigma$) \\
Mesitylene (C\textsubscript{9}H\textsubscript{12}) & 0.59 (\textless 2$\sigma$) \\
1-Nitropropane (C\textsubscript{3}H\textsubscript{7}NO\textsubscript{2}) & 0.58 (\textless 2$\sigma$) \\
Pentane (C\textsubscript{5}H\textsubscript{12}) & 0.58 (\textless 2$\sigma$) \\
1,2,3-Trimethylbenzene (C\textsubscript{9}H\textsubscript{12}) & 0.55 (\textless 2$\sigma$) \\
2,4,4-Trimethyl-2-pentene (C\textsubscript{8}H\textsubscript{16}) & 0.54 (\textless 2$\sigma$) \\
DL-Limonene (C\textsubscript{10}H\textsubscript{16}) & 0.54 (\textless 2$\sigma$) \\
Allylamine (C\textsubscript{3}H\textsubscript{7}N) & 0.53 (\textless 2$\sigma$) \\
Diethylformamide (C\textsubscript{5}H\textsubscript{11}NO) & 0.53 (\textless 2$\sigma$) \\
Myrcene (C\textsubscript{10}H\textsubscript{16}) & 0.53 (\textless 2$\sigma$) \\
Piperidine (C\textsubscript{5}H\textsubscript{11}N) & 0.53 (\textless 2$\sigma$) \\
D-Limonene (C\textsubscript{10}H\textsubscript{16}) & 0.51 (\textless 2$\sigma$) \\
Nicotine (C\textsubscript{10}H\textsubscript{14}N\textsubscript{2}) & 0.51 (\textless 2$\sigma$) \\
Cadaverine (C\textsubscript{5}H\textsubscript{14}N\textsubscript{2}) & 0.50 (\textless 2$\sigma$) \\
2-Methyl-1-pentene (C\textsubscript{6}H\textsubscript{12}) & 0.49 (\textless 2$\sigma$) \\
Iodomethane (CH\textsubscript{3}I) & 0.49 (\textless 2$\sigma$) \\
N,N-Diethylaniline (C\textsubscript{10}H\textsubscript{15}N) & 0.49 (\textless 2$\sigma$) \\
Propane (C\textsubscript{3}H\textsubscript{8}) & 0.49 (\textless 2$\sigma$) \\
Methylisocyanate (C\textsubscript{2}H\textsubscript{3}NO) & 0.48 (\textless 2$\sigma$) \\
Diethyl sulfide (C\textsubscript{4}H\textsubscript{10}S) & 0.47 (\textless 2$\sigma$) \\
2,4,4-Trimethyl-1-pentene (C\textsubscript{8}H\textsubscript{16}) & 0.46 (\textless 2$\sigma$) \\
3-Methylpentane (C\textsubscript{6}H\textsubscript{14}) & 0.46 (\textless 2$\sigma$) \\
4-Methyl-1-pentene (C\textsubscript{6}H\textsubscript{12}) & 0.46 (\textless 2$\sigma$) \\
1-Butene (C\textsubscript{4}H\textsubscript{8}) & 0.44 (\textless 2$\sigma$) \\
2-Picoline (C\textsubscript{6}H\textsubscript{7}N) & 0.44 (\textless 2$\sigma$) \\
Chloromethane (CH\textsubscript{3}Cl) & 0.43 (\textless 2$\sigma$) \\
Methyl isothiocyanate (C\textsubscript{2}H\textsubscript{3}NS) & 0.43 (\textless 2$\sigma$) \\
tert-Butylbenzene (C\textsubscript{10}H\textsubscript{14}) & 0.43 (\textless 2$\sigma$) \\
Cyclooctane (C\textsubscript{8}H\textsubscript{16}) & 0.42 (\textless 2$\sigma$) \\
2-Butene (C\textsubscript{4}H\textsubscript{8}) & 0.41 (\textless 2$\sigma$) \\
Isoprene (C\textsubscript{5}H\textsubscript{8}) & 0.40 (\textless 2$\sigma$) \\
o-Xylene (C\textsubscript{8}H\textsubscript{10}) & 0.40 (\textless 2$\sigma$) \\
Hexadecane (C\textsubscript{16}H\textsubscript{34}) & 0.39 (\textless 2$\sigma$) \\
3-Methylhexane (C\textsubscript{7}H\textsubscript{16}) & 0.38 (\textless 2$\sigma$) \\
Isopentane (C\textsubscript{5}H\textsubscript{12}) & 0.38 (\textless 2$\sigma$) \\
Tetrahydrothiophene (C\textsubscript{4}H\textsubscript{8}S) & 0.38 (\textless 2$\sigma$) \\
n-Butylamine (C\textsubscript{4}H\textsubscript{11}N) & 0.37 (\textless 2$\sigma$) \\
1-Octene (C\textsubscript{8}H\textsubscript{16}) & 0.36 (\textless 2$\sigma$) \\
cis-4-Methyl-2-pentene (C\textsubscript{6}H\textsubscript{12}) & 0.36 (\textless 2$\sigma$) \\
Methylacrylonitrile (C\textsubscript{4}H\textsubscript{5}N) & 0.36 (\textless 2$\sigma$) \\
Chloroethane (C\textsubscript{2}H\textsubscript{5}Cl) & 0.35 (\textless 2$\sigma$) \\
Sulfur monoxyde (SO) & 0.35 (\textless 2$\sigma$) \\
Butane (C\textsubscript{4}H\textsubscript{10}) & 0.34 (\textless 2$\sigma$) \\
Thiophosphoryl chloride (SPCl\textsubscript{3}) & 0.34 (\textless 2$\sigma$) \\
Ethyl benzene (C\textsubscript{8}H\textsubscript{10}) & 0.32 (\textless 2$\sigma$) \\
n-Undecane (C\textsubscript{11}H\textsubscript{24}) & 0.32 (\textless 2$\sigma$) \\
Phosphorus mononitride (PN) & 0.32 (\textless 2$\sigma$) \\
Acetone cyanohydrin (C\textsubscript{4}H\textsubscript{7}NO) & 0.31 (\textless 2$\sigma$) \\
Pentadecane (C\textsubscript{15}H\textsubscript{32}) & 0.31 (\textless 2$\sigma$) \\
Ethylene sulfide (C\textsubscript{2}H\textsubscript{4}S) & 0.30 (\textless 2$\sigma$) \\
Hexane (C\textsubscript{6}H\textsubscript{14}) & 0.30 (\textless 2$\sigma$) \\
Isocumene (C\textsubscript{9}H\textsubscript{12}) & 0.30 (\textless 2$\sigma$) \\
Nitroethane (C\textsubscript{2}H\textsubscript{5}NO\textsubscript{2}) & 0.30 (\textless 2$\sigma$) \\
Triethylamine (C\textsubscript{6}H\textsubscript{15}N) & 0.29 (\textless 2$\sigma$) \\
Nitrobenzene (C\textsubscript{6}H\textsubscript{5}NO\textsubscript{2}) & 0.28 (\textless 2$\sigma$) \\
Vinyl toluene (C\textsubscript{9}H\textsubscript{10}) & 0.28 (\textless 2$\sigma$) \\
1-Propanethiol (C\textsubscript{3}H\textsubscript{8}S) & 0.27 (\textless 2$\sigma$) \\
1-Undecene (C\textsubscript{11}H\textsubscript{22}) & 0.27 (\textless 2$\sigma$) \\
Cyclopropane (C\textsubscript{3}H\textsubscript{6}) & 0.27 (\textless 2$\sigma$) \\
Pyridine (C\textsubscript{5}H\textsubscript{5}N) & 0.27 (\textless 2$\sigma$) \\
2-Nitropropane (C\textsubscript{3}H\textsubscript{7}NO\textsubscript{2}) & 0.26 (\textless 2$\sigma$) \\
Methanesulfonyl-chloride (CH\textsubscript{3}ClO\textsubscript{2}S) & 0.26 (\textless 2$\sigma$) \\
Allyl chloride (C\textsubscript{3}H\textsubscript{5}Cl) & 0.25 (\textless 2$\sigma$) \\
Isobutene (C\textsubscript{4}H\textsubscript{8}) & 0.25 (\textless 2$\sigma$) \\
22-Dimethylbutane (C\textsubscript{6}H\textsubscript{14}) & 0.23 (\textless 2$\sigma$) \\
Ethyl-mercaptan (C\textsubscript{2}H\textsubscript{6}S) & 0.22 (\textless 2$\sigma$) \\
Aniline (C\textsubscript{6}H\textsubscript{7}N) & 0.21 (\textless 2$\sigma$) \\
Isohexane (C\textsubscript{6}H\textsubscript{14}) & 0.21 (\textless 2$\sigma$) \\
p-Xylene (C\textsubscript{8}H\textsubscript{10}) & 0.21 (\textless 2$\sigma$) \\
Styrene (C\textsubscript{8}H\textsubscript{8}) & 0.21 (\textless 2$\sigma$) \\
Quinoline (C\textsubscript{9}H\textsubscript{7}N) & 0.20 (\textless 2$\sigma$) \\
sec-Amylamine (C\textsubscript{5}H\textsubscript{13}N) & 0.20 (\textless 2$\sigma$) \\
4-Ethyltoluene (C\textsubscript{9}H\textsubscript{12}) & 0.19 (\textless 2$\sigma$) \\
2-Ethyltoluene (C\textsubscript{9}H\textsubscript{12}) & 0.18 (\textless 2$\sigma$) \\
2-Methyl-2-pentene (C\textsubscript{6}H\textsubscript{12}) & 0.18 (\textless 2$\sigma$) \\
2,4-Dimethylpentane (C\textsubscript{7}H\textsubscript{16}) & 0.18 (\textless 2$\sigma$) \\
Nitric oxide (NO) & 0.17 (\textless 2$\sigma$) \\
Silicon monoxyde (SiO) & 0.17 (\textless 2$\sigma$) \\
Sulfur mononitride (NS) & 0.17 (\textless 2$\sigma$) \\
Sulfuryl fluoride (SO\textsubscript{2}F\textsubscript{2}) & 0.17 (\textless 2$\sigma$) \\
Thiophene (C\textsubscript{4}H\textsubscript{4}S) & 0.17 (\textless 2$\sigma$) \\
1-Decene (C\textsubscript{10}H\textsubscript{20}) & 0.16 (\textless 2$\sigma$) \\
2-Methyl-1-propanethiol (C\textsubscript{4}H\textsubscript{10}S) & 0.16 (\textless 2$\sigma$) \\
Ethyl nitrite (C\textsubscript{2}H\textsubscript{5}NO\textsubscript{2}) & 0.16 (\textless 2$\sigma$) \\
n-Decane (C\textsubscript{10}H\textsubscript{22}) & 0.16 (\textless 2$\sigma$) \\
Sulfuryl-chloride (SO\textsubscript{2}Cl\textsubscript{2}) & 0.16 (\textless 2$\sigma$) \\
2-Propanethiol (C\textsubscript{3}H\textsubscript{8}S) & 0.14 (\textless 2$\sigma$) \\
4-Picoline (C\textsubscript{6}H\textsubscript{7}N) & 0.14 (\textless 2$\sigma$) \\
Morpholine (C\textsubscript{4}H\textsubscript{9}NO) & 0.14 (\textless 2$\sigma$) \\
Isobutane (C\textsubscript{4}H\textsubscript{10}) & 0.13 (\textless 2$\sigma$) \\
Isopropylamine (C\textsubscript{3}H\textsubscript{9}N) & 0.13 (\textless 2$\sigma$) \\
Methyl nitrite (CH\textsubscript{3}NO\textsubscript{2}) & 0.12 (\textless 2$\sigma$) \\
Octane (C\textsubscript{8}H\textsubscript{18}) & 0.12 (\textless 2$\sigma$) \\
1,2,3,5-Tetramethylbenzene (C\textsubscript{10}H\textsubscript{14}) & 0.11 (\textless 2$\sigma$) \\
Hydrogen sulfide (H\textsubscript{2}S) & 0.11 (\textless 2$\sigma$) \\
Benzenethiol (C\textsubscript{6}H\textsubscript{6}S) & 0.10 (\textless 2$\sigma$) \\
tert-Butylmercaptan (C\textsubscript{4}H\textsubscript{10}S) & 0.10 (\textless 2$\sigma$) \\
Phosphine (PH\textsubscript{3}) & 0.09 (\textless 2$\sigma$) \\
Acetonitrile (C\textsubscript{2}H\textsubscript{3}N) & 0.08 (\textless 2$\sigma$) \\
Benzonitrile (C\textsubscript{7}H\textsubscript{5}N) & 0.08 (\textless 2$\sigma$) \\
Nitromethane (CH\textsubscript{3}NO\textsubscript{2}) & 0.06 (\textless 2$\sigma$) \\
Fluoride (HF) & 0.05 (\textless 2$\sigma$) \\
Sulfur hexafluoride (SF\textsubscript{6}) & 0.05 (\textless 2$\sigma$) \\
2,3-Dimethylbutane (C\textsubscript{6}H\textsubscript{14}) & 0.04 (\textless 2$\sigma$) \\
Nitrogen trifluoride (NF\textsubscript{3}) & 0.04 (\textless 2$\sigma$) \\
1,3-Butadiene (C\textsubscript{4}H\textsubscript{6}) & 0.03 (\textless 2$\sigma$) \\
n-Nonane (C\textsubscript{9}H\textsubscript{20}) & 0.03 (\textless 2$\sigma$) \\
1-Nonene (C\textsubscript{9}H\textsubscript{18}) & 0.02 (\textless 2$\sigma$) \\
2,6-Diethylaniline (C\textsubscript{10}H\textsubscript{15}N) & 0.01 (\textless 2$\sigma$) \\
Nitrous oxide (N\textsubscript{2}O) & 0.01 (\textless 2$\sigma$) \\
Isooctane (C\textsubscript{8}H\textsubscript{18}) & 0.00 (\textless 2$\sigma$) \\
Perchloromethyl mercaptan (CCl\textsubscript{4}S) & 0.00 (\textless 2$\sigma$) \\
\enddata\end{deluxetable*}